\documentclass[useAMS,usenatbib]{mn2e}
\usepackage{epsfig}
\usepackage{amsmath}

\def\lsim{~\rlap{$<$}{\lower 1.0ex\hbox{$\sim$}}}

\def\gsim{~\rlap{$>$}{\lower 1.0ex\hbox{$\sim$}}}

\newcommand{\ud}{\,\mathrm{d}}

\newcommand{\xvec}{\boldsymbol{x}}
\newcommand{\yvec}{\boldsymbol{y}}
\newcommand{\zvec}{\boldsymbol{z}}

\newcommand{\rvec}{\boldsymbol{r}}

\newcommand{\kvec}{\boldsymbol{k}}

\newcommand{\Uvec}{\boldsymbol{U}}

\newcommand{\Upsilonvec}{\boldsymbol{\Upsilon}}

\def \HII{H\,{\scriptsize II}\,\,}

\def \HIeqn{\rm H\,{\scriptscriptstyle I}}

\title[Polarised foreground removal using RM synthesis]{Polarised foreground removal at low radio frequencies using rotation measure synthesis: uncovering the signature of hydrogen reionisation}

\author[Geil, Gaensler \& Wyithe]{Paul M. Geil\thanks{Email: pmgeil@unimelb.edu.au}, J. Stuart B. Wyithe\\
School of Physics, The University of Melbourne, Parkville, Victoria, Australia}

\author[Geil, Gaensler \& Wyithe]{Paul M Geil$^1$\thanks{Email: pmgeil@unimelb.edu.au}, B M Gaensler$^{2,3}$, J. Stuart B. Wyithe$^{1,3}$\\
$^1$School of Physics, The University of Melbourne, Parkville, Victoria, 3010, Australia\\
$^2$School of Physics, The University of Sydney, New South Wales, 2006, Australia\\
$^3$ARC Centre of Excellence for All-sky Astrophysics (CAASTRO)\\}

\begin{document}

\maketitle

\label{firstpage}
\begin{abstract}

Measurement of redshifted 21-cm emission from neutral hydrogen promises to be the most effective method for studying the reionisation history of hydrogen and, indirectly, the first galaxies. These studies will be limited not by raw sensitivity to the signal, but rather, by bright foreground radiation from Galactic and extragalactic radio sources and the Galactic continuum. In addition, leakage due to gain errors and non-ideal feeds conspire to further contaminate low-frequency radio obsevations. This leakage leads to a portion of the complex linear polarisation signal finding its way into Stokes~$I$, and inhibits the detection of the non-polarised cosmological signal from the epoch of reionisation. In this work, we show that rotation measure synthesis can be used to recover the signature of cosmic hydrogen reionisation in the presence of contamination by polarised foregrounds. To achieve this, we apply the rotation measure synthesis technique to the Stokes~$I$ component of a synthetic data cube containing Galactic foreground emission, the effect of instrumental polarisation leakage, and redshifted 21-cm emission by neutral hydrogen from the epoch of reionisation. This produces an effective Stokes~$I$ Faraday dispersion function for each line of sight, from which instrumental polarisation leakage can be fitted and subtracted. Our results show that it is possible to recover the signature of reionisation in its late stages ($z \approx 7$) by way of the 21-cm power spectrum, as well as through tomographic imaging of ionised cavities in the intergalactic medium.

\end{abstract}


\section{Introduction}
\label{Introduction}

Statistical observations of the redshifted 21-cm emission from neutral hydrogen during the epoch of reionisation (EoR) promise to provide a wealth of information about the properties of neutral hydrogen at high redshift, as well as some of the fundamental astrophysics behind the reionisation process and the first luminous objects. While density perturbations in the matter distribution mediate fluctuations in the 21-cm signal both prior to and following reionisation, during the reionisation era the relation between the 21-cm power spectrum and the underlying matter power spectrum is complex and, in its late stages, dominated by the formation of large ionised ``bubbles" \citep{furl2004b,mcquinn2006}. These bubbles of ionised hydrogen imprint features on the 21-cm power spectrum that reflect the luminosity and clustering of ionising sources responsible for reionisation. The 21-cm power spectrum can be used as a statistical test to distinguish candidate reionisation models, as well as to constrain the history and morphology of reionisation \citep{barkana2009}.

Measuring this weak cosmic signal will be challenging however, owing to contamination from a large number of astrophysical and non-astrophysical components. Indeed, astrophysical foregrounds are typically brighter than the cosmological 21-cm signal by 4--5 orders of magnitude \citep[see, e.g.,][]{dimatteo2002,oh2003}. Some of these foregrounds also have bright polarised counterparts. As a result, significant contamination will occur in the non-polarised (Stokes~$I$) component of observations due to non-ideal feed configuration and electronic errors in interferometers. This process is known as instrumental polarisation leakage. Various experiments are planned to measure 21-cm emission from the neutral intergalactic medium (IGM), including the Low Frequency Array\footnote{http://www.lofar.org/} (LOFAR), the Murchison Widefield Array\footnote{http://www.mwatelescope.org/} (MWA) and the Precision Array to Probe Epoch of Reionization\footnote{http://astro.berkeley.edu/$\sim$dbacker/eor/} (PAPER). These instruments will be capable of operating at full-Stokes, meaning they will be capable of separating non-polarised signals from polarised signal.

It has been suggested that foregrounds can be removed through continuum subtraction \citep[e.g.,][]{gnedin2004,wang2006}, or using the differences in symmetry from power spectra analysis \citep{morales2004,zaldarriaga2004}. This has been shown in recent work \citep{jelic2008,geil2008b} to be an effective method for the purpose of detecting the EoR signal. On the other hand, eliminating contamination of Stokes~$I$ due to instrumental polarisation leakage has not been treated. This paper demonstrates a method that removes polarised contamination by utilising a procedure known as rotation measure (RM) synthesis.

Rotation measure synthesis exploits the Fourier relationship between the complex linear polarisation signal, $P(\lambda^2)$, and the Faraday dispersion function, $F(\phi)$, making it possible to identify components of polarised signal with specific Faraday depth, $\phi$ \citep[see, e.g.,][]{burn1966,brentjens2005}. We modify this technique so as to make it applicable to the non-polarised Stokes~$I$ signal, calculating an effective Stokes~$I$ Faraday dispersion function, $F_I(\phi)$. The three-dimensional Stokes~$I$ data set is first re-sampled along the frequency axis into the natural $\lambda^2$-basis for Faraday rotated emission. A one-dimensional Fourier transform is then applied along each line of sight. Any feature observed in $F(\phi)$ with a non-zero Faraday depth will, by the definition of our model, be polarised contamination, thereby making it possible to identify peaks of leaked polarised signal in $F_I(\phi)$. An iterative cleaning process, together with knowledge of the instrument's specific leakage behaviour, can be applied to $F_I(\phi)$ in order to attempt to subtract this contamination. An inverse one-dimensional Fourier transform is then applied along each line of sight, followed by a coordinate transformation back to frequency space. Applying this procedure to all lines of sight in a field of view produces a three-dimensional Faraday dispersion cube; hence the use of the term Faraday tomography.

This paper is organised as follows: Section~\ref{Astrophysical sources} discusses the astrophysical sources we model in this work. Section~\ref{Basics} discusses the basic principles of radio polarimetry. Section~\ref{Astrophysical and instrumental effects} discusses the astrophysical and instrumental effects we simulate in this work. The generation of synthetic Stokes~$I$ observation data cubes is outlined in Section~\ref{Synthetic data cube}. Section~\ref{Foreground subtraction} explains the continuum foreground removal strategy and introduces the leaked polarised foreground component subtraction algorithm we have developed. This section also includes a summary of the assumptions used throughout. In Section~\ref{Spherically averaged power spectra}, we discuss the use of spherically averaged power spectra as a key statistical measure of cosmic reionisation, and our method of computing statistical instrumental sensitivity. Section~\ref{Results} presents the results of our simulations and their analyses. Section~\ref{Summary} is a summary of this paper. 

Throughout this paper we adopt a set of cosmological parameters consistent with the \textit{Wilkinson Microwave Anisotropy Probe} (\textit{WMAP}) \citep{komatsu2009} for a flat $\Lambda$CDM universe: $\Omega_{\rm m} = 0.27$ (dark matter and baryons); $\Omega_{\Lambda} = 0.73$ (cosmological constant); $\Omega_{\rm b} = 0.046$ (baryons); $h = 0.7$ (Hubble parameter); $n_{\rm s} = 1$ (primordial spectrum index); $\sigma_8 = 0.8$ (primordial spectrum normalisation). All distances are in comoving units unless stated otherwise.

\section{Astrophysical sources}
\label{Astrophysical sources}

In this work, we model two astrophysical sources: the non-polarised 21-cm emission from the epoch of reionisation and diffuse Galactic synchrotron emission (DGSE). We consider the total intensity of DGSE which appears in the non-polarised Stokes~$I$ parameter, as well as the its polarised component.

\subsection{Epoch of reionisation signal}
\label{Epoch of reionisation signal}

Reionisation starts with isolated regions of ionised hydrogen (H\,{\scriptsize II}) forming around galaxies and quasars, which later grow and merge to surround clusters of galaxies. The reionisation process is complete when these regions overlap and fill the volume between galaxies. Stars are expected to be the main ionising sources during the EoR \citep[see, e.g.,][]{srbin2007,faucher2008,loeb2009}, although other sources such as quasars have also been suggested. To which level each of these contribute is still an open question. The stellar contribution begins with the formation of massive, metal-free stars that form in the shallow potential wells of the first collapsed dark matter halos ($z \sim 15$), and continues by way of a normal population of less massive stars that form from the metal-enriched gas in larger dark matter halos present at lower redshift ($z \sim 6$). We use a semi-numerical model for the reionisation of a three-dimensional volume of the IGM by galaxies. This model is described very briefly below and we direct the reader to \cite{geil2008a} for further details.

Briefly, we begin by simulating the linear matter overdensity field $\delta(\xvec,z) \equiv \rho_{\rm m}(\xvec,z)/\bar{\rho}_{\rm m}-1$ inside a periodic, comoving, cubic region of volume $V = L^{3}$, by calculating the density contrast in Fourier space $\hat{\delta}(\kvec,z)$ corresponding to a $\Lambda$CDM power spectrum \citep{eh1999} extrapolated to a specified redshift $z$. The semi-analytic model used to compute the relation between the local dark matter overdensity and the ionisation state of the IGM is based on the model described by \cite{wl2007} and \cite{wm2007}. The sum of astrophysical effects included in our model leads to the growth of \HII regions via a phase of percolation during which individual \HII regions overlap around clustered sources in overdense regions of the universe. Where necessary, we consider a model in which the mean IGM is fully reionised by redshift $z_{\rm ov} = 6$ \citep{fan2006,gnedin2006,white2003}.

\begin{figure*}
\begin{center}
\includegraphics[width=17.5cm]{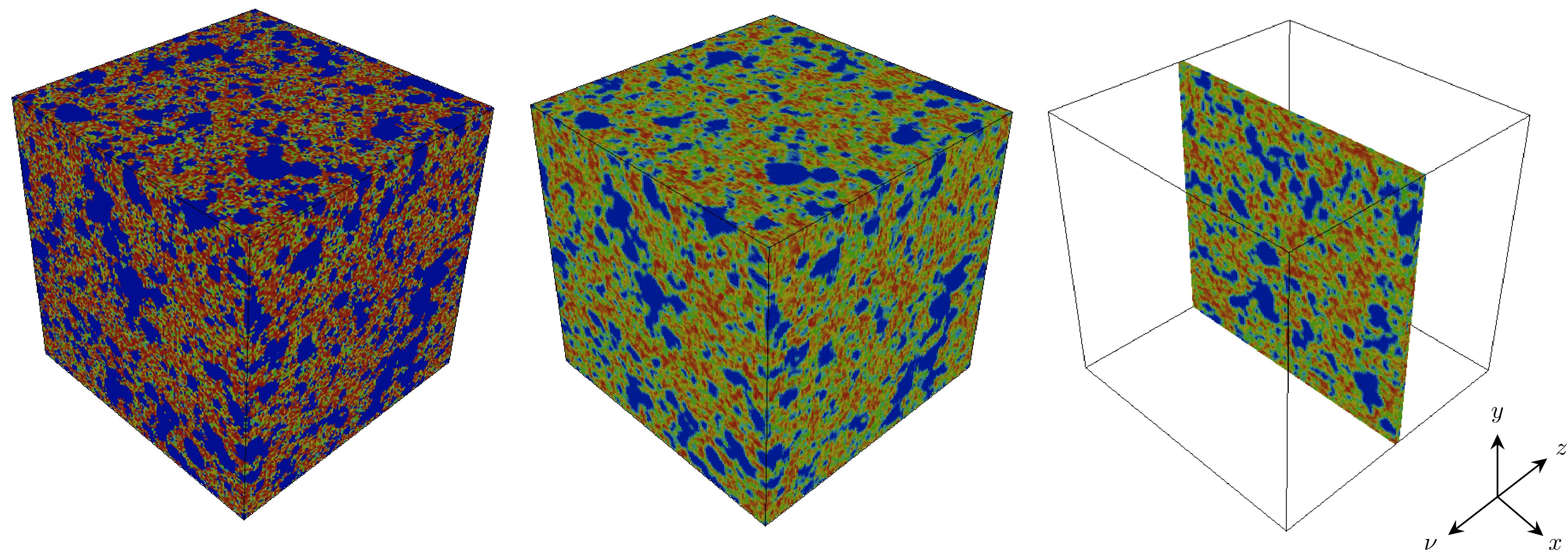} 
\caption{Three-dimensional model epoch of reionisation signal at $z = 7$ showing fully ionised regions with no 21-cm emission (\textit{blue}, 0~mK) and partially ionised, 21-cm signal emitting regions (\textit{red}, 11~mK). Each box has a comoving side length of approximately 524~Mpc. \textit{Left:} Non-instrumentally smoothed. \textit{Centre:} Instrumentally smoothed in the sky-plane using a Guassian tapered synthesised beam with a 5~arc minute full-width half-maximum. \textit{Right:} Instrumentally smoothed, central frequency channel slice.}
\label{eor_model}
\end{center}
\end{figure*}

The corresponding average 21-cm brightness temperature contrast of a gas parcel with the cosmic microwave background (CMB) \citep[][]{madau1997} is given by
\begin{eqnarray}
\delta T_{\rm b} &=& 26~x_{\HIeqn} (1+\delta_b) \left(1 - \frac{T_{\gamma}}{T_{\rm s}}\right) \left( \frac{\Omega_b h^2}{0.022} \right)  \nonumber \\
&\times&\left( \frac{0.15}{\Omega_m h^2} \frac{1+z}{10} \right)^{1/2}~{\rm mK}.
\end{eqnarray}
Here, the spin temperature, $T_{\rm s}$, CMB brightness temperature, $T_{\gamma}$, baryonic overdensity, $\delta_b$, (which we assume to be equal to the dark matter overdensity) and neutral fraction, $x_{H\,{\scriptsize I}\,\,}$, are calculated for a location $\xvec$ at redshift $z$.  We assume $T_{\rm s} \gg T_{\gamma}$ during the epoch of reionisation \citep{ciardi2003,furl2006} and ignore the enhancement of brightness temperature fluctuations due to peculiar velocities in overdense regions \citep{bharadwaj2005,bl2005}. Peculiar velocities were included in the semi-numerical model of \cite{mesinger2007}, who found their effect to be small on scales $\sim 10$~Mpc. Figure~\ref{eor_model} shows an example of a synthetic three-dimensional epoch of reionisation signal data cube at $z = 7$ with and without the effect of limited instrumental angular resolution (see Section~\ref{The synthesised beam and beam depolarisation} for a description of the synthesised beam resolution).

For the purpose of demonstrating foreground removal, we do not consider the evolution of the IGM in the line-of-sight direction within the observed bandpass. Therefore, the realisation of the density and ionisation state of the IGM in each frequency channel is considered to be at the same stage of cosmic evolution. This is a valid approximation when the range in redshift corresponding to the simulated bandwidth is small compared with the central redshift of the bandwidth $\Delta z/z \ll 1$ for redshifts greater than the redshift of overlap $z > z_{\rm ov}$. We also assume that the reionisation signal is non-polarised and is therefore present in Stokes~$I$ only.

\subsection{Diffuse Galactic synchrotron emission}
\label{Diffuse Galactic synchrotron emission}

Diffuse Galactic synchrotron emission originates from relativistic electrons in the interstellar medium interacting with the Galactic magnetic field. Although we are primarily concerned with the polarised component in this paper, we discuss our model for total intensity of synchrotron emission here because of the proportionality between the two, as well as to demonstrate the combined effect of continuum and polarised-component foreground removal. For more detailed non-polarised and polarised foreground models, including synchrotron emission from discrete sources such as supernova remnants and free-free emission from diffuse ionised gas, see \cite{jelic2008,bowman2009,jelic2010}

\subsubsection{Total intensity of diffuse Galactic synchrotron emission}
\label{Non-polarised DGSE}

Foreground contamination and its removal could have significant consequences for the detectability of the weak, redshifted 21-cm signal. Indeed, the total intensity of foregrounds will be brighter than the cosmological 21-cm signal by 4--5 orders of magnitude. The three main sources of foreground contamination of the 21-cm signal are DGSE (which comprises $\sim$~70~per~cent near 150~MHz), extragalactic point sources ($\sim$~27~per~cent) and Galactic bremsstrahlung ($\sim$~1~per~cent) \citep{shaver1999}. The frequency dependence of each of these foregrounds can be approximated by a power law with a running spectral index \citep{shaver1999,tegmark2000}. While the sum of power laws is not in general a power law, over a relatively narrow frequency range, $B$, (such as that considered in this paper where $B/\nu\ll 1$), a Taylor expansion around a power law can be used to describe the spectral shape. We therefore also approximate the sum of foregrounds as a power law with a running spectral index, and specialise to the case of Galactic synchrotron emission, which dominates the foregrounds. For the purpose of demonstrating the removal of polarised foreground leakage, we assume that the brightest point sources have already been removed.

\begin{figure*}
\begin{center}
\includegraphics[width=15.5cm]{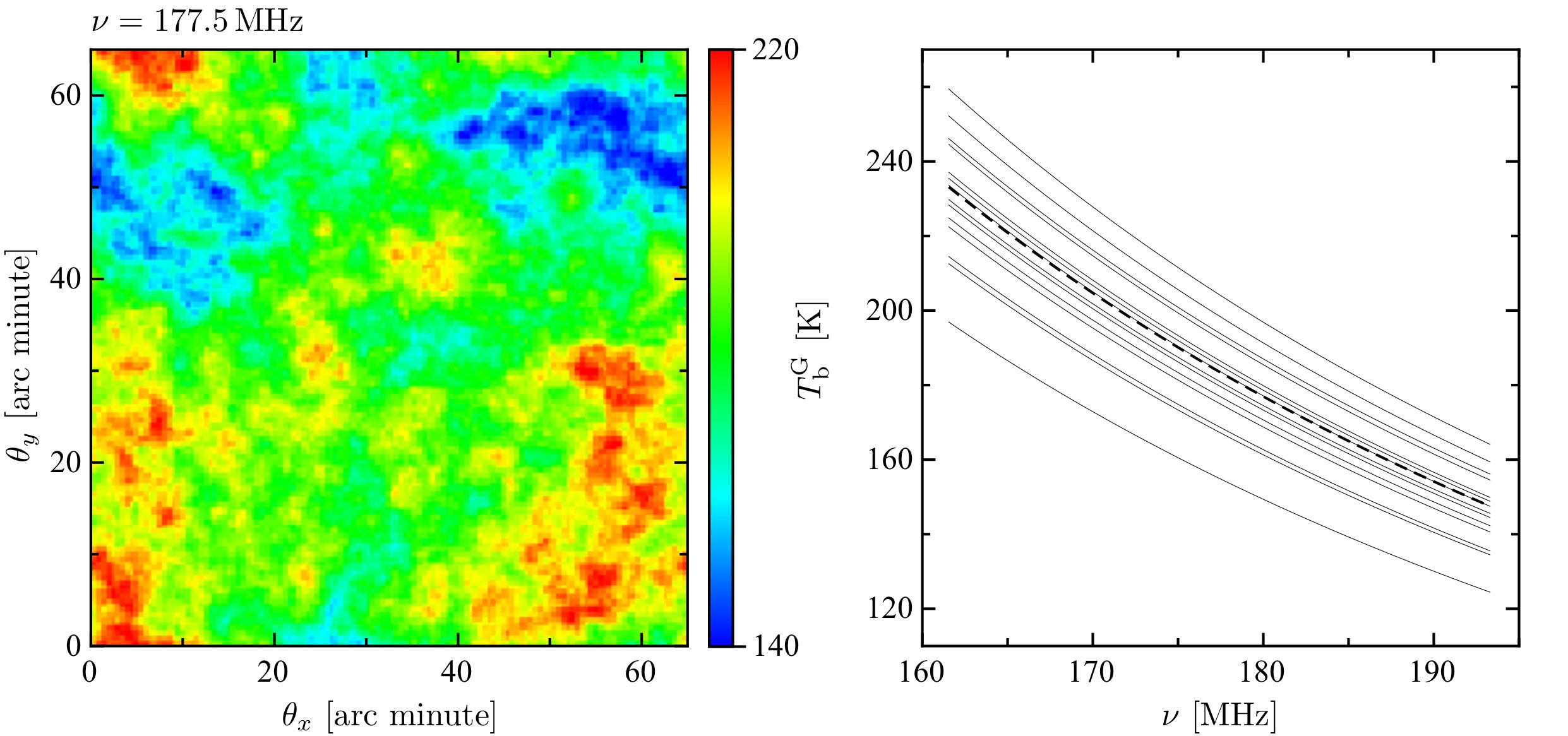} 
\caption{\textit{Left:} A single frequency slice at $\nu = 177.5$~MHz showing the angular structure of the diffuse Galactic synchrotron foreground emission, for which $(\bar{T}^{\rm G}_{\rm b},\sigma_{T}) \approx (183,16)$~K. \textit{Right:} DGSE as a function of frequency for a number of random lines of sight. The dashed line shows the value of $T^{\rm G}_{\rm b}(\nu)$ for $\delta T_{\rm b,0}^{\rm G} = 0$.}
\label{dgse_map}
\end{center}
\end{figure*}

The total intensity of Galactic synchrotron emission varies as a function of both sky position and frequency. We model the frequency and angular dependence of Galactic synchrotron foreground emission \citep[following][]{shaver1999,tegmark2000,wang2006} by first constructing a realisation of the angular fluctuations in the foreground, $\delta T^{\rm G}_{\rm b,0}(\vec{\theta})$, at a particular frequency within the band of interest using a two-dimensional Gaussian random field. We then add a mean sky brightness, $\bar{T}^{\rm G}_{\rm b,0}$, to our two-dimensional realisation of brightness temperature fluctuations. This foreground plane is then extended into three dimensions by extrapolation, using the running power law form along each line of sight,
\begin{eqnarray}
\label{TGb}
T^{\rm G}_{\rm b}(\vec{\theta},\nu) &=& \left[\bar{T}^{\rm G}_{\rm b,0} + \delta T^{\rm G}_{\rm b,0}(\vec{\theta})
\right]\nonumber\\
& & \times \left( \frac{\nu}{\nu_0}\right)^{-\alpha_{\rm syn}-\Delta \alpha_{\rm syn} \rm log_{10}\left(\frac{\nu}{\nu_0}\right)}.
\end{eqnarray}
We have modelled the sky-plane variation of $\alpha_{\rm syn}$ in this paper by drawing a value for $\Delta \alpha_{\rm syn}$ from a Gaussian probability distribution with a standard deviation of 0.1, for every line of sight. Note that over a data cube at the frequencies of interest, $\bar{T}_{\rm b}^{\rm G} \sim 200$~K. The strategy for removing continuum foreground contamination is discussed in Section~\ref{Continuum component subtraction}. Figure~\ref{dgse_map} includes a single frequency slice at $\nu = 177.5$~MHz (\textit{left}), showing the angular structure of the total intensity of diffuse Galactic synchrotron foreground emission, for which $(\bar{T}^{\rm G}_{\rm b},\sigma_{T}) = (183,16)$~K. Also shown (\textit{right}) is the DGSE as a function of frequency for a number of lines of sight. The dashed line shows the value of $T^{\rm G}_{\rm b}(\nu)$ for $\delta T_{\rm b,0}^{\rm G} = 0$.

Although the mean brightness temperature of the DGSE is $\sim 200$~K, the interferometric response to this emission will only be the fluctuation from the mean brightness temperature within the primary beam\footnote{The primary beam is the field of view of a single antenna tile. MWA antennas have fields of view of $\sim 240\lambda^2$~square degrees (to the first null), where $\lambda$ is in the range $1 < \lambda < 4$~m.}. This is because the signal recieved by any single antenna does not contribute to the synthesised signal, i.e., the instrument as a whole responds to non-zero baseline spacing only. Such fluctuations are generally much smaller than the mean over the primary beam. We retain the primary beam's mean brightness temperature in our DGSE model with no consequence, since any constant signal along a line of sight is removed during the continuum foreground subtraction stage.

\subsubsection{Polarised DGSE}
\label{Polarised DGSE}

The polarisation state of a stream of radiation may be specified by the four Stokes parameters $I$, $Q$, $U$ and $V$. Linearly polarised radiation can be specified using $Q$ and $U$ only and is often represented by the complex linear polarisation $P$, given by
\begin{eqnarray}
\label{eq:P}
P = Q + iU = pI\exp(2i\chi),
\end{eqnarray}
where $Q = \Re[P]$ and $U = \Im[P]$, $p$ is the intrinsic degree of polarisation and $\chi$ is the position angle (or polarisation angle) of the linearly polarised radiation. These quantities are related by
\begin{eqnarray}
\label{eq:p and chi}
pI = \sqrt{Q^2 + U^2},\qquad \chi = \frac{1}{2}\arctan\left(\frac{U}{Q}\right).
\end{eqnarray}

We model the polarised component of the DGSE as being proportional to its total intensity, $T_{\rm b}^{\rm G}(\nu)$, such that
\begin{eqnarray}
\label{QandU}
Q_{\rm G}(\nu) &=& pT_{\rm b}^{\rm G}(\nu)\cos(2\chi),\\
U_{\rm G}(\nu) &=& p T_{\rm b}^{\rm G}(\nu)\sin(2\chi),
\end{eqnarray}
where $T_{\rm b}^{\rm G}(\nu)$ is the total Galactic synchrotron brightness temperature, $\chi$ is the polarisation angle (line-of-sight-dependent) and the polarisation fraction, $p$, is dependent on the brightness temperature spectral index, $\alpha_{\rm syn}$, through
\begin{eqnarray}
p = \frac{3\alpha_{\rm syn} - 3}{3\alpha_{\rm syn} - 1} \approx 0.7,
\end{eqnarray}
\citep{laroux1961}.

\cite{bernardi2009,bernardi2010} give constraints on the level of polarised DGSE using Westerbork Synthesis Radio Telescope data at 150~MHz and arcmin scales. They found that the rms value for polarised foreground is $\approx 7.2$~K on 4~arc minute scales. Using a polarisation fraction of 0.7, the fluctuations in our polarised foreground at the same frequency with a 5~arc minute full width half maximum beam are of the same order of magnitude ($0.7 \times 25~{\rm K} = 18 ~{\rm K}$).

\section{Basics of radio polarimetry}
\label{Basics}

\subsection{Faraday rotation}
\label{Faraday rotation}

The rotation of the plane of polarisation of radiation (specified by $\chi$) propagating through a magneto-ionic medium results from the birefringence of the plasma. This effect, known as Faraday rotation, has a nonlinear dependence on frequency such that
\begin{eqnarray}
\label{eq:chi}
\chi = \chi_0 + \phi\lambda^2,
\end{eqnarray}
where $\phi$ is the Faraday depth and $\chi_0$ is the non-rotated polarisation angle. Faraday depth is defined by
\begin{eqnarray}
\phi = 8.1 \times 10^5\, \int_L^0 n_e \textit{\textbf{B}} \cdot \ud\textit{\textbf{r}}~{\rm rad~m}^{-2},
\end{eqnarray}
where $n_e$ is the electron density in cm$^{-3}$, $\textit{\textbf{B}}$ is the magnetic field in Gauss, $L$ is the distance between the point of emission and the observer in parsec and $\ud\textit{\textbf{r}}$ is an infinitesimal path vector in parsec (\textit{away} from the point of emission). A positive Faraday depth indicates an integrated magnetic field pointing toward the observer, however, the Faraday depth may be negative due to the integrated orientation of the magnetic field pointing away from the observer along the line of sight.

Along any particular line of sight, there may be any number of Faraday screens at different Faraday depths by which polarised radiation may be Faraday rotated having been emitted with a unique initial polarisation angle. We include only one screen per line of sight in our simulations in order to show the principle behind the instrumental polarisation leakage cleaning process.

There are two planned target EoR fields for the MWA, centred on $(\alpha,\delta) = (4^{\rm h},\,-30^{\circ})$ and (7$^{\rm h}40^{\rm m},10^{\circ})$, which correspond to the Galactic coordinates $(l,b) \approx (228^{\circ},-49^{\circ})$ and $(209^{\circ},+15^{\circ})$ respectively. The interpolated RM values (and thus Faraday depth) of both of these fields including extragalactic sources are $20 \lsim \,\, \phi \lsim \,\, 50$~rad\,m$^{-2}$ \citep{johnston-hollitt2004}. It is reasonable to assume that the magnitude of Faraday depth is either less than or not much larger than extragalactic RM values in the same direction. RM ranges in other directions have been given by \cite{haverkorn2003} ($-17 \lsim \,\, \phi \lsim \,\, 10$~rad\,m$^{-2}$) and \cite{bernardi2010} ($|\phi|\,\lsim\,\,10$~rad\,m$^{-2}$). In our synthetic data cube, we have chosen to include a single Galactic Faraday screen with $1 \lsim \,\, \phi \lsim \,\, 5$~rad\,m$^{-2}$ in every line of sight. This range of shallow Faraday depth has been chosen despite the larger observed values since we have found  contamination by polarised signal passing through screens at small Faraday depth more troublesome to clean. Therefore, our results represent a worst case scenario. We model the angular distribution of Faraday depth of the Galactic screen by a small-gradient-plane over the sky-plane (such that, for generality, $\nabla \phi(x,y) \propto \hat{\xvec} + \hat{\yvec}$). This is done to approximate the small, slowly varying Faraday depth over the proposed EoR fields of the MWA.

\subsection{Faraday dispersion function}
\label{Faraday dispersion function}

In dealing with the RM synthesis of a distribution of synchrotron-emitting and Faraday rotating regions along a line of sight, we follow the work of \cite{brentjens2005} and direct readers to their paper for a more thorough explanation, as well as \cite{schnitzeler2009} for an application of the technique.

\subsubsection{Continuous case}

The Faraday dispersion function, $F(\phi)$, can be defined through
\begin{eqnarray}
P(\lambda^2) = \int F(\phi)\exp(2i\phi\lambda^2)\,d\phi,
\end{eqnarray}
where the frequency dependence of $P$ has been given in terms of $\lambda^2$ since $\lambda^2$ and $\phi$ form a Fourier-variable pair. In practice, the sampling of $\lambda^2$ will be limited by the finite frequency bandpass of the correlator. Therefore, we introduce a weighting or sampling function, $W(\lambda^2)$, such that
\begin{eqnarray}
\label{eq:Pobs}
P_{\rm obs}(\lambda^2) = W(\lambda^2) \int F(\phi)\exp(2i\phi\lambda^2)\,d\phi,
\end{eqnarray}
where, for example,
\begin{eqnarray}
W(\lambda^2) = \left\{
\begin{array}{rl}
1 & \text{if } {\rm channel\,\,is\,\,recorded}\\
0 & \text{if } {\rm otherwise.}
\end{array} \right.
\end{eqnarray}
Other forms of the sampling function may include non-uniform weighting to account for non-ideal response for certain channels (at the shoulders of the bandpass, for example). Equation~(\ref{eq:Pobs}) can be Fourier inverted to find $F(\phi)$ and convolved with the inverse Fourier transform of the sampling function. This gives the Faraday dispersion function that would be obtained given the limited sampling of $P$,
\begin{eqnarray}
\label{eq:Fobs}
F_{\rm obs}(\phi) &\equiv& F(\phi) \ast R(\phi)\nonumber \\
&=& K \int P_{\rm obs}(\lambda^2) \exp(-2i\phi\lambda^2)\,d\lambda^2,
\end{eqnarray}
where the rotation measure spread function, $R(\phi)$, is defined by
\begin{eqnarray}
\label{eq:RMSF}
R(\phi) = K \int W(\lambda^2) \exp(-2i\phi\lambda^2)\,d\lambda^2,
\end{eqnarray}
and the appropriate normalisation factor is
\begin{eqnarray}
\label{eq:K}
K = \left[ \int W(\lambda^2)\,d\lambda^2 \right]^{-1}.
\end{eqnarray}

\begin{figure}
\begin{center}
\includegraphics[width=8.5cm]{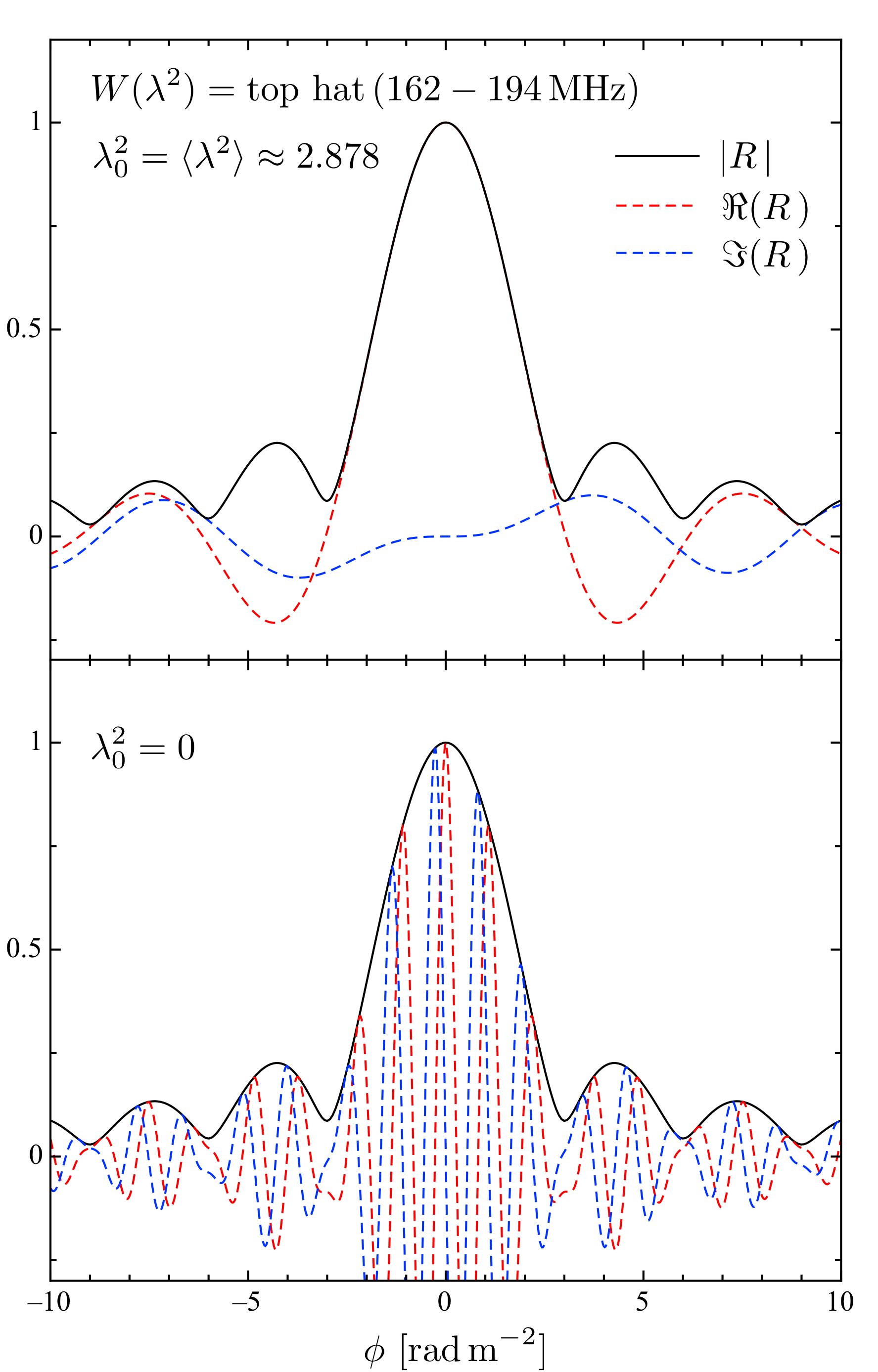} 
\caption{Demonstration of requirements for back rotation of polarisation angle showing the magnitude (\textit{solid black}), real part (\textit{dashed red}) and imaginary part (\textit{dashed blue}) of $R(\phi)$ for a 32~MHz bandpass of uniform response centred on 177.5~MHz. \textit{Bottom}: Polarisation angle derotated using $\lambda_0^2 = 0$. \textit{Top}: Polarisation angle derotated using the weighted mean value of $\lambda^2$.}
\label{fig:RMSF}
\end{center}
\end{figure}

The rotation measure spread function is a complex-valued function whose magnitude is peaked and symmetric about $\phi = 0$. The lower panel of Figure~\ref{fig:RMSF} shows the magnitude, real part and imaginary part of $R(\phi)$ for a 32~MHz bandpass of uniform response centred on 177.5~MHz. It can be seen that both the real and imaginary parts oscillate rapidly throughout the central response peak. This makes it difficult to determine $\chi$, which given the form of Equations~(\ref{eq:Fobs}) and (\ref{eq:RMSF}) is the polarisation angle derotated back to $\lambda^2 = 0$. It is possible, instead, to shift the $\lambda^2$ variable by some amount, affecting the oscillitory behaviour of the complex components of $R(\phi)$. In general, the optimal shift is given by the weighted mean value of $\lambda^2$,
\begin{eqnarray}
\label{eq:lambda_0_squared}
\lambda_0^2 = \int W(\lambda^2)\lambda^2\,d\lambda^2 \Big/ \int W(\lambda^2)\,d\lambda^2,
\end{eqnarray}
It is simple to show, as can be seen in the top panel of Figure~\ref{fig:RMSF}, that this forces $\partial {\Im[R(\phi)]}/\partial\phi |_{\phi = 0} = 0$, thereby making the evaluation of the polarisation angle using Equation~(\ref{eq:p and chi}) more accurate. Hence, using Equation~(\ref{eq:chi}), we can now write the derotated polarisation angle as
\begin{eqnarray}
\chi_0 = \chi(\lambda_0^2) - \phi\lambda_0^2.
\end{eqnarray}

Dropping the `obs' subscript hereafter and making the shift in $\lambda^2$, we have
\begin{eqnarray}
P(\lambda^2) &=& W(\lambda^2) \int F(\phi)\exp\left[2i\phi(\lambda^2-\lambda_0^2)\right]\,d\phi,\\
\label{eq:F}
F(\phi) &=& K \int P(\lambda^2) \exp\left[-2i\phi(\lambda^2-\lambda_0^2)\right]\,d\lambda^2,\\
\label{eq:R}
R(\phi) &=& K \int W(\lambda^2) \exp\left[-2i\phi(\lambda^2-\lambda_0^2)\right]\,d\lambda^2,
\end{eqnarray}
where $K$ is defined as before in Equation~(\ref{eq:K}).

\subsubsection{Discrete case}

Since a correlator samples signals discretely, we must use discrete rather than continuous Fourier transformations. To do this, consider a correlator which provides $N_{\rm ch}$ frequency channels over a bandpass of width $B$ centred on $\nu_0$, such that each channel has a width $\Delta\nu = B/N_{\rm ch}$. The correlator gives signal measurements for equally spaced channels in frequency, however, the parameter we wish to work with is $\lambda^2$. We make the following approximations for the channel centres and widths. The central $\lambda^2$ of channel $j$ is given by
\begin{eqnarray}
\label{eq:lambda2c}
\lambda_j^2 &=& \int_{\Delta\nu_j} \lambda^2\,d\nu \Big/ \int_{\Delta\nu_j} d\nu\nonumber \\
&=& \frac{c^2}{\nu_j^2}\left[1+\frac{1}{4}\left(\frac{\Delta\nu}{\nu_j}\right)^2\right] + \mathcal{O}\left[\left(\frac{\Delta\nu}{\nu_j}\right)^3\right],
\end{eqnarray}
where $\nu_j$ is the central frequency of the channel. The width of channel $j$ is given by
\begin{eqnarray}
\Delta\lambda_j^2 &=& \frac{2c^2\Delta\nu}{\nu_j^3}\left[1+\frac{1}{2}\left(\frac{\Delta\nu}{\nu_j}\right)^2\right] + \mathcal{O}\left[\left(\frac{\Delta\nu}{\nu_j}\right)^4\right].
\end{eqnarray}
If $\phi\Delta\lambda^2 \ll 1$ for all channels, Equations~(\ref{eq:K}), (\ref{eq:F}) and (\ref{eq:R}) can be approximated by the following Fourier sums:
\begin{eqnarray}
\label{eq:Fsum}
F(\phi) &\approx& K \sum_{j=0}^{N_{\rm ch}-1} P_j \exp\left[-2i\phi(\lambda_j^2-\lambda_0^2)\right] \Delta\lambda_j^2;\\
R(\phi) &\approx& K \sum_{j=0}^{N_{\rm ch}-1} w_j \exp\left[-2i\phi(\lambda_j^2-\lambda_0^2)\right] \Delta\lambda_j^2;\\
K &\approx& \left( \sum_{j=0}^{N_{\rm ch}-1} w_j \right)^{-1},
\end{eqnarray}
where $\lambda_j^2$ is the central $\lambda^2$ of channel $j$ as approximated by Equation~(\ref{eq:lambda2c}).

In the lag-domain, the correlator we consider corresponds to $N_{\rm ch}$ lag channels covering a total time $\Delta t = 1/\Delta\nu$, with lag spacing $\delta t = 1/B$. By choosing to perform the frequency-lag conversion with respect to the central frequency, the recorded lag spectrum will be centred on $t = 0$. This enables the recovery of Faraday depths of either sign in the Faraday dispersion function.

\subsection{Equivalent Faraday dispersion function for Stokes~$I$}
\label{Faraday dispersion function for Stokes I}

Unlike the complex linear polarisation, $P$, the non-polarised Stokes~$I$ signal is a real-valued function. Despite this, it is possible to calculate an effective Stokes~$I$ Faraday dispersion function for this signal by assigning its imaginary part to zero in all channels. Then, in analogy with Equation~(\ref{eq:Fsum}),
\begin{eqnarray}
F_I(\phi) \approx K \sum_{j=0}^{N_{\rm ch}-1} I_j \exp\left[-2i\phi(\lambda_j^2-\lambda_0^2)\right] \Delta\lambda_j^2,
\end{eqnarray}
where $I_j$ is the Stokes~$I$ signal measurement for channel $j$. Since the Stokes~$I$ response is a real function, by the property of the Fourier transform, the effective Stokes~$I$ Faraday dispersion function is therefore Hermitian, i.e., $F^*_I(\phi) = F_I(-\phi)$. Because of this redundancy, we may plot $F_I(\phi)$ for $\phi \geq 0$ only without any loss of information.

\section{Astrophysical and instrumental effects}
\label{Astrophysical and instrumental effects}

In this section, we discuss some of the astrophysical and instrumental effects relevant to radio polarimetry and radio interferometry in general, and introduce the effect of instrumental polarisation leakage. For convenience, we consider the propagation of the Stokes vector visibility throughout the system in terms of the brightness temperature.

Throughout, we consider a radio interferometer with specifications comparable with the MWA. The MWA will consist of $\approx 500$ tiles, each with 16 cross-dipoles with 1.07~m spacing and an effective area $A_{\rm e} \approx 16(\lambda^2/4)$ ($A_{\rm e}$ is capped for $\lambda > 2.1$~m). For simplicity, we model the beam-formed response of each tile as that of a single cross-dipole. We assume a continuous antenna density profile, $\rho_{\rm a} \propto r^{-2}$, within a 750~m radius with a constant core density of approximately one tile per 36~m$^2$. Section~\ref{Fiducial array configuration} of the Appendix gives more detail on the fiducial array configuration considered.

\subsection{Bandwidth depolarisation}
\label{Bandwidth depolarisation}

\begin{figure}
\begin{center}
\includegraphics[width=8.5cm]{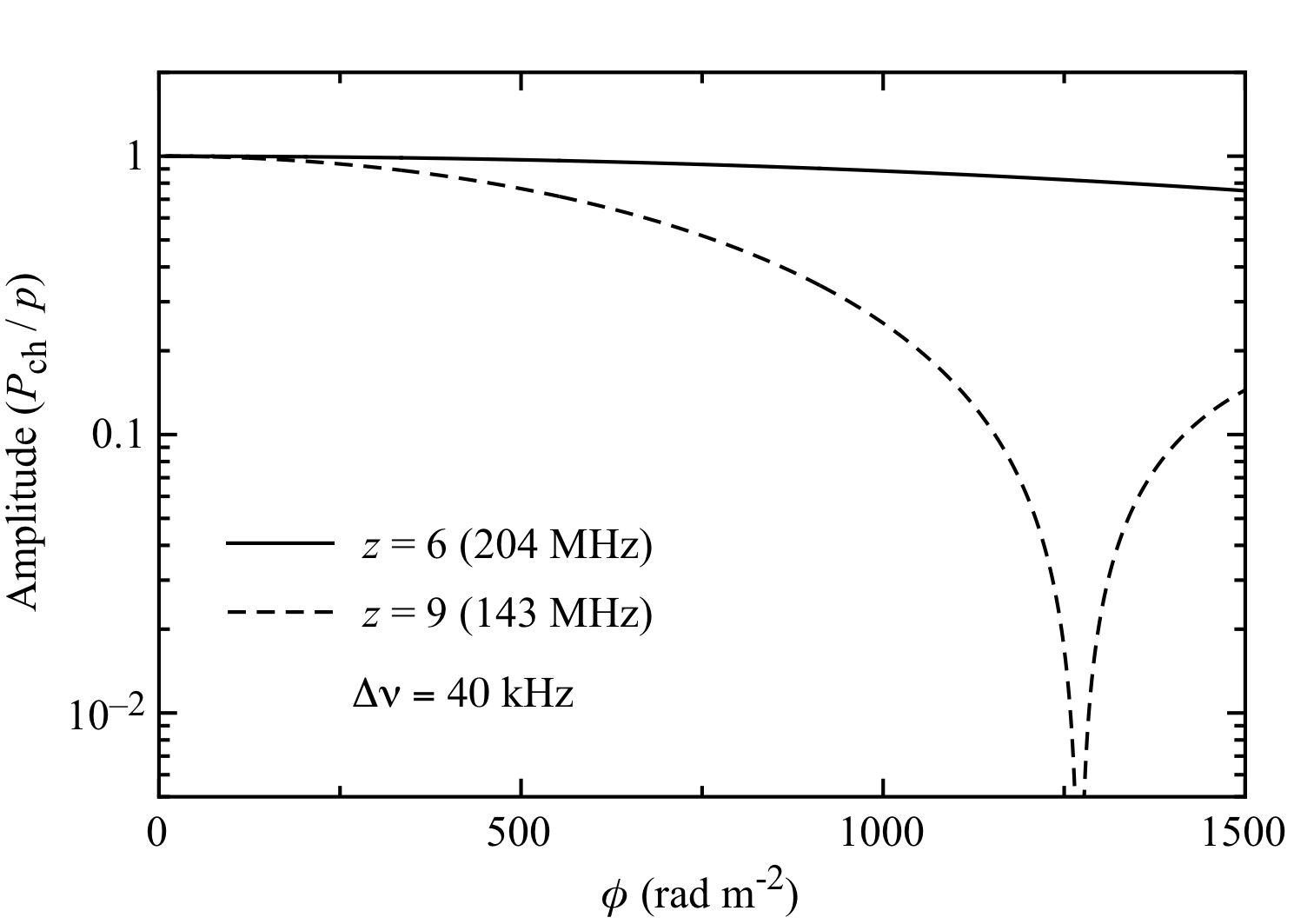} 
\caption{Bandwidth depolarisation for the MWA [$\Delta\nu = 40$~kHz centred on $\nu_{\rm c} = \nu_{21}/(1+z)$].}
\label{MWAdepol}
\end{center}
\end{figure}

The integrated complex polarisation over each frequency channel will suffer from depolarisation due to the deconstructive summation of linear polarisation vectors within the small but finite range of frequencies in each channel. The integrated complex polarisation for a channel with frequency range $\nu_1 \leq \nu \leq \nu_2$ is given by
\begin{eqnarray}
P_{\rm ch}(\nu_1,\nu_2) = \int_{\nu_1}^{\nu_2} P(\nu)\,d\nu \Big/ \int_{\nu_1}^{\nu_2} d\nu.
\end{eqnarray}
Given a channel of width $\Delta\nu = \nu_2 - \nu_1$ centred on $\nu_{\rm c} = (\nu_1 + \nu_2)/2$, such that $\Delta\nu/\nu_{\rm c} \ll 1$,
\begin{eqnarray}
\label{Pch}
P_{\rm ch}(\nu) = p\frac{\sin\alpha}{\alpha}e^{2i\phi\lambda_{\rm c}^2},
\end{eqnarray}
where $\alpha = 2\phi\lambda_{\rm c}^2\Delta\nu/\nu_{\rm c}$, and it is assumed $p$ is independent of frequency. Figure~\ref{MWAdepol} shows the bandwidth depolarisation for the MWA centred on $\nu = 143$~MHz and 204~MHz (corresponding to the redshifted 21-cm emission frequency for $z = 9$ and 6 respectively). We include the effects of bandwidth depolarisation by generating over-sampled synthetic Stokes~$Q$ and $U$ spectra and then summing the contributions over each channel to produce the final spectra with the desired number of channels.

Note that the detectable range and sampling resolution of Faraday depth can be found in the following way. The sinc-factor effectively enveloping $P_{\rm ch}$ vanishs when $\alpha = \pi/2$. Solving for $\phi$ when $P_{\rm ch} = 0$ gives
\begin{eqnarray}
\phi_{\rm max} = \frac{\pi\nu_0 t_{\rm max}}{2\lambda^2_0} = \frac{\pi\nu_0}{4\lambda^2_0\Delta\nu},
\end{eqnarray}
where $\lambda^2_0$ is the corresponding central value of $\lambda^2$ given by Equation~(\ref{eq:lambda_0_squared}). From this, the Faraday depth sampling interval is
\begin{eqnarray}
\Delta\phi \equiv \frac{2\phi_{\rm max}}{N_{\rm ch}} = \frac{\pi\nu_0}{2\lambda^2_0 B}.
\end{eqnarray}
Note that the central lag channel (if $N_{\rm ch}$ is odd, otherwise the central two channels) contains all signals with $| \phi | < \Delta\phi$.

\subsection{The synthesised beam and beam depolarisation}
\label{The synthesised beam and beam depolarisation}

Due to the finite length of the longest baseline, radio interferometers have a correspondingly finite angular resolution. This can be quantified in terms of the synthesised beamwidth, given by
\begin{eqnarray}
\Delta\theta_{\rm b} = \frac{0.705\lambda}{D},
\end{eqnarray}
where $\lambda$ is the wavelength of observing and $D$ is the maximum baseline length\footnote{Note that the beamwidth referred to by many authors is $\Delta\theta = 1.22\lambda/D \sim \lambda/D$, which is the highest angular resolution meeting the Rayleigh criterion. This corresponds to the central peak to first null separation in the resulting beam-response pattern of a circular aperture. Other conventions include: full-beamwidth between first nulls, $\Delta\theta_{\rm FWFN} = 2.44\lambda/D$; full-beamwidth at half-power, $\Delta\theta_{\rm FWHP} = 1.02\lambda/D$; and full-beamwidth at half-maximum, $\Delta\theta_{\rm FWHM} = 0.705\lambda/D$. We have adopted the last of these conventions in this paper.}. When calculating the actual angular resolution achievable in the field of observation, the projection of the baselines onto the sky-plane must be taken into consideration. Throughout this work we assume observations of a zenith-field only.

In addition to the limited angular resolution imposed by the maximum baseline length of the interferometer, sky-plane images can  be smoothed by applying a weighted filter to the two-dimensional Fourier transform of each image for every frequency channel. This can be used to take weight from data that have been poorly sampled due to insufficient baseline coverage for example, at the expense of angular resolution. The particular type of filter used depends on the final observation product required. For all 21-cm simulations in this work we have applied a Gaussian filter in Fourier-space which corresponds in image-space to a Gaussian smoothing function of width of 5~arc minutes for all frequency channels. 

In addition to bandwidth depolarisation, a second cause of depolarisation is beam averaging. Unresolved polarised sources within a synthesised beam will add vectorally and may therefore superimpose destructively. Similarly, a spatial gradient in Faraday depth over the synthesised beam will produce a variation in the angle of polarisation as a function of position. If this variation is signiffcant when averaged across a single synthesised beam, the polarised fraction can be reduced. For simplicity, we have not modelled these effects in this work.

\subsection{Representation}
\label{Representation}

In this section, we closely follow the notation of \cite{hamaker1996}, \cite{sault1996} and \cite{tms}. 

The electric field of the most general homogeneous plane wave propagating in the direction $\hat{\kvec}$, recieved at the feed position of antenna `$i$' can be given with respect to the Cartesian coordinates $x$, $y$ and $z$ by
\begin{eqnarray}
\textbf{e}_i(t) = \left(
\begin{array}{c}
E_{i,x}\\
E_{i,y}
\end{array}\right)\exp[-i\omega t],
\end{eqnarray}
where the amplitudes $E_{i,x}$ and $E_{i,y}$ are complex numbers, allowing for a possible phase difference. If these components have the same phase the wave is said to be \textit{linearly polarised}; if different it is said to be \textit{elliptically polarised}, with the special case of a $\pi/2$ phase difference giving a \textit{circularly polarised} wave.

The plane waves recieved by the antennas may have been modified by a number of spurious effects that occur along the path between the source and the feeds. The resulting electrical signal is also modified by a number of intentional and other spurious effects that occur along the path between the feeds and the correlator. These effects include: physical effects such as the Faraday rotation by magneto-ionic media between the signal source and the feeds; geometrical effects such as the rotation of the feeds with respect to the polarisation coordinates of the signal, including the projection of the dipoles onto the observed sky plane; instrumental effects such as the conversion from an electric field to a voltage in the feed electronics; signal amplification; and the effects of polarisation leakage or variation of gain with frequency. These effects can be described by linear transformations and represented by what are known as Jones operators \citep{jones1941}, $\textbf{J}$, transforming as
\begin{eqnarray}
\textbf{e}_i^{\prime} = \textbf{J}_i\,\textbf{e}_i,
\end{eqnarray}
where the prime indicates the vector after transformation.

The combined effect of these transformations can be mathematically described by a transformation made up of the product of any number of Jones matrices. For simplicity, we only consider Faraday rotation, complex gain and polarisation leakage in this work, assuming that all other effects have been accounted for, by calibration for example. Therefore, the combined Jones matrix for interferometer arm `$i$' is given by
\begin{eqnarray}
\textbf{J}_i = \textbf{G}_i\,\textbf{D}_i\,\textbf{F}_i,
\end{eqnarray}
where (in order of operation) $\textbf{F}_i$ is the Faraday rotation operator, $\textbf{D}_i$ is the operator that accounts for polarisation leakage through deviations of the actual feed from the ideal and $\textbf{G}_i$ is the complex receiver gain operator. These operators are given by
\begin{eqnarray}
\textbf{F}_i = \left(
\begin{array}{cc}
\cos \Delta\chi_i & -\sin \Delta\chi_i\\
\sin \Delta\chi_i & \cos \Delta\chi_i
\end{array}\right),
\end{eqnarray}
\begin{eqnarray}
\textbf{D}_i = \left(
\begin{array}{cc}
1 & d_{i,xy}\\
-d_{i,yx} & 1
\end{array}\right),
\end{eqnarray}
\begin{eqnarray}
\textbf{G}_i = \left(
\begin{array}{cc}
g_{i,x} & 0\\
0 & g_{i,y}
\end{array}\right),
\end{eqnarray}
where $\Delta\chi_i$ is the angle of Faraday rotation, $d_{i,xy}$ is the spurious sensitivity of the $x$ receptor to the $y$ polarisation and vice versa, and $g_{i,x}$ is the gain error coefficient for the $x$ receptor. Note that $\textbf{F}_i$ is real-valued whereas $\textbf{D}_i$ and $\textbf{G}_i$ are complex-valued.

For a single interferometric baseline, the measured visibilities can be described by a $2 \times 2$ coherency matrix\footnote{Throughout this paper, vectors and operators in the coherencey or Stokes representation are denoted by column vectors (lower-case bold non-italic) and matrices (upper-case bold non-italic) respectively, in order to distinguish them from vectors with no specific basis (bold italic).}, $\textbf{C}_{ij}$, which is formed by the time-averaged outer product of the two-dimensional, complex vector input signals $\textbf{e}_i$ and $\textbf{e}_j$ from the two antennas making up the baseline,
\begin{eqnarray}
\textbf{C}_{ij} = \langle\,\textbf{e}_i(t) \otimes \textbf{e}_j^{\dagger}(t)\,\rangle,
\end{eqnarray}
where the dagger symbol represents the adjoint operation of complex conjugation and transposition. It is also possible to describe the measured visibilities by a four-dimensional coherency column vector,
\begin{eqnarray}
\textbf{c}_{ij} &=& \langle\,\textbf{e}_i(t) \otimes \textbf{e}_j^{*}(t)\,\rangle\\
&=& \left(
\begin{array}{c}
E_{i,x} E_{j,x}^*\\
E_{i,x} E_{j,y}^*\\
E_{i,y} E_{j,x}^*\\
E_{i,y} E_{j,y}^*
\end{array}\right).
\end{eqnarray}
Including the effect of the transformations, we have
\begin{eqnarray}
\textbf{c}_{ij}^{\prime} &=& \langle\,\textbf{e}_i^{\prime}(t) \otimes \textbf{e}_j^{\prime *}(t)\,\rangle \nonumber\\
&=& \langle\,\textbf{J}_i\,\textbf{e}_i(t) \otimes [\textbf{J}_j\,\textbf{e}_j(t)]^*\,\rangle \nonumber\\
&=& \langle\,\textbf{J}_i\,\textbf{e}_i(t) \otimes \textbf{J}_j^*\,\textbf{e}_j^*(t)\,\rangle \nonumber\\
&=& (\textbf{J}_i \otimes \textbf{J}_j^*)\langle\,\textbf{e}_i(t) \otimes \textbf{e}_j^*(t)\,\rangle  \nonumber\\
&=& \textbf{J}_{ij}\textbf{c}_{ij},
\end{eqnarray}
where $\textbf{J}_{ij} \equiv \textbf{J}_i \otimes \textbf{J}_j^*$ and we have made use of the following identity for the outer product:
\begin{eqnarray}
(\textbf{A} \textbf{B}) \otimes (\textbf{C} \textbf{D}) = (\textbf{A} \otimes \textbf{C})(\textbf{B} \otimes \textbf{D}).
\end{eqnarray}

As the Stokes parameters are used almost universally in radio polarimetry, we now make the conversion from the coherency representation to the Stokes visibility representation. A distinction should be made at this point: the Stokes \textit{parameters}, $I$, $Q$, $U$ and $V$, are used when describing the intensity or brightness of radiation, whereas the Stokes \textit{visibilities} are used to represent the complex visibilities as measured by an interferometer. The context in which these terms are used provides the means for disambiguation. The Stokes visibility vector as measured by a single baseline made up of antennas $i$ and $j$, $\textbf{s}_{ij} = [\,I_{ij}\;Q_{ij}\;U_{ij}\;V_{ij}\,]^{\rm T}$, is related to the coherency vector via a simple coordinate transformation, $\textbf{S}$, through
\begin{eqnarray}
\textbf{s}_{ij} = \textbf{S}\,\textbf{c}_{ij},
\end{eqnarray}
where
\begin{eqnarray}
\textbf{S} = \left(
\begin{array}{cccc}
1 & 0 & 0 & 1\\
1 & 0 & 0 & -1\\
0 & 1 & 1 & 0\\
0 & -i & i & 0
\end{array}\right),
\end{eqnarray}
the inverse of which is
\begin{eqnarray}
\textbf{S}^{-1} = \frac{1}{2}\left(
\begin{array}{cccc}
1 & 1 & 0 & 0\\
0 & 0 & 1 & i\\
0 & 0 & 1 & -i\\
1 & -1 & 0 & 0
\end{array}\right).
\end{eqnarray}
Note that the complex vector input signals and Jones matrices are two-dimensional whereas the coherence vectors and Stokes visibility vectors are four-dimensional. 

Operators that act in coherency- or Stokes-space can be transformed between representations using the same transformation matrix, $\textbf{S}$, used for the coherency and Stokes vectors, through
\begin{eqnarray}
\label{eq:op_xfm}
\textbf{A}^{\rm (S)} = \textbf{S}\,\textbf{A}^{\rm (C)}\,\textbf{S}^{-1},
\end{eqnarray}
where $\textbf{A}^{\rm (C)}$ is the operator in coherency representation and $\textbf{A}^{\rm (S)}$ is the operator in Stokes representation (operators in the Stokes representation that are used to reproduce the effect of one or more non-depolarising, transforming elements are called Mueller matrices). Therefore, we can write the full Mueller matrix for the single baseline system as
\begin{eqnarray}
\label{eq:JStokes}
\textbf{M}_{ij} = \textbf{S}\,\textbf{J}_{ij}\,\textbf{S}^{-1} = \textbf{S}\,(\textbf{J}_i \otimes \textbf{J}^*_j)\,\textbf{S}^{-1}.
\end{eqnarray}

\subsection{Thin Faraday screens}
\label{Thin Faraday screens}

Assuming the signal through both interferometic arms experience the same degree of Faraday rotation $\Delta\chi$ (a valid assumption for baselines shorter than the scale of fluctuations in the ionosphere), applying the coherency--Stokes visibility transformation on the Faraday rotation operators of a single baseline gives
\begin{eqnarray}
\textbf{M}_{\rm F} =
\left(\begin{array}{cccc}
1 & 0 & 0 & 0\\
0 & \cos(2\Delta\chi) & -\sin(2\Delta\chi) & 0\\
0 & \sin(2\Delta\chi) & \cos(2\Delta\chi) & 0\\
0 & 0 & 0 & 1
\end{array}\right).
\end{eqnarray}
This is equivalent to a rotation of the $P = [Q,U]^{\rm T}$ vector through an angle of $2\Delta\chi$ radian. So we see that Faraday rotation in the reduced Stokes representation ($Q$ and $U$ only) is the same as that in the complex vector representation (where $\textbf{F}$ acts). This is reconcilable with Equation~(\ref{eq:P}) since $P$ is rotated by $2\Delta\chi = 2\phi\lambda^2$ from an initial angle $2\chi_0$ resulting in a final angle of $2\chi = 2\chi_0 + 2\Delta\chi$.

The efffect of thin Faraday screens on incident synchrotron emission can be modelled by
\begin{eqnarray}
\label{Q_F}
Q_{\rm F}(\lambda_j^2) &=& \sum_{k=1}^{N_{\rm F}} p_k(\lambda_j^2)\cos(2\chi_{jk}),\\
\label{U_F}
U_{\rm F}(\lambda_j^2) &=& \sum_{k=1}^{N_{\rm F}} p_k(\lambda_j^2)\sin(2\chi_{jk}),
\end{eqnarray}
where $N_{\rm F}$ is the number of thin Faraday screens and $\chi_{jk} = (\chi_0)_k + \phi_k(\lambda_j^2-\lambda_0^2)$, $(\chi_0)_k$ being the intrinsic polarisation angle at the $k$th screen.

\subsection{Instrumental polarisation leakage}
\label{leakage}

In practice, feeds are non-ideal and this leads to leakage between all Stokes parameters, $I,\,Q,\,U$ and $V$. The treatment of instrumental polarisation leakage is similar to that as for Faraday rotation, but complicated by the fact that in general the feed errors corrupt all four Stokes parameters and cannot be assumed to be the same in all antennas.

\begin{figure}
\begin{center}
\includegraphics[width=8.5cm]{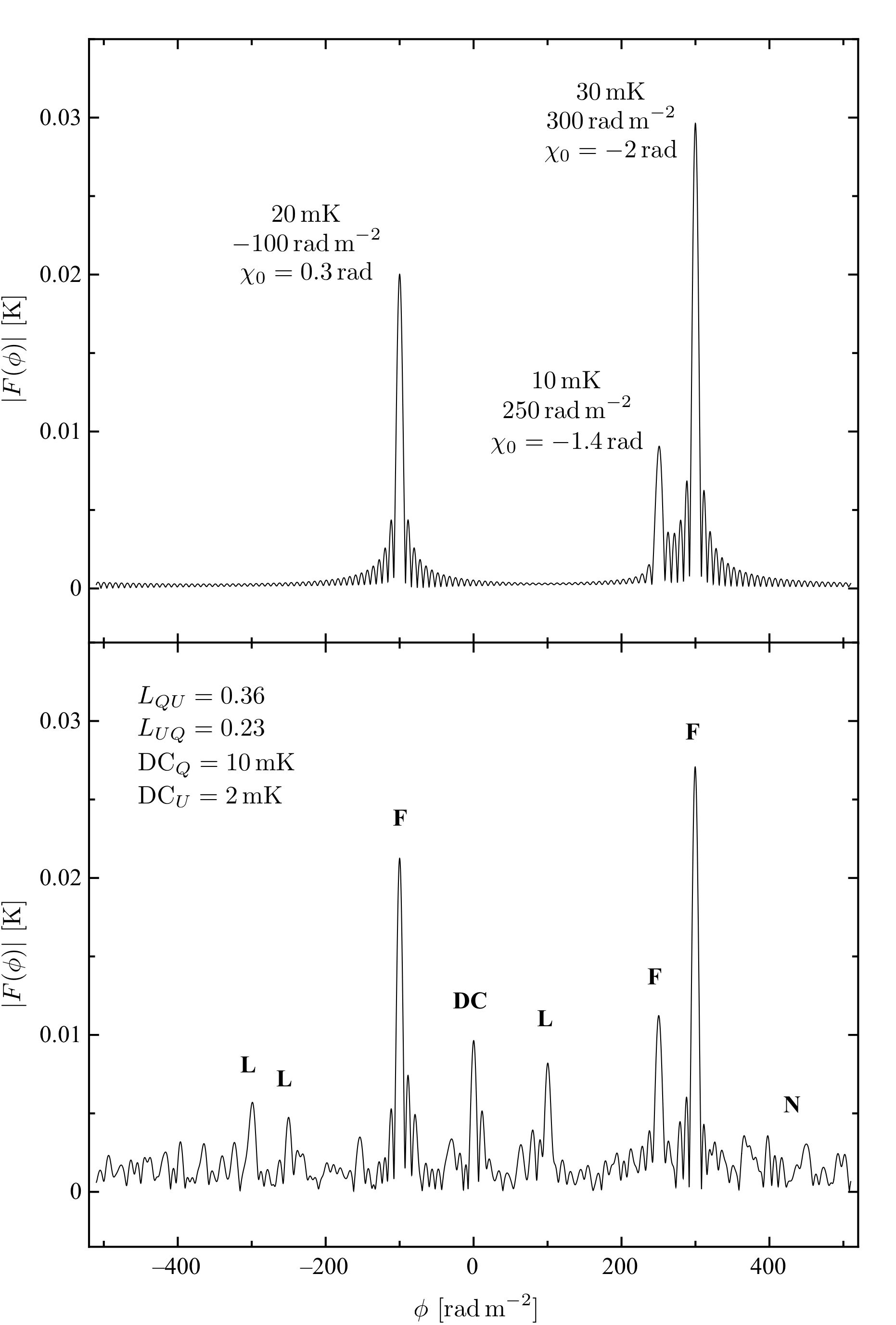} 
\caption{Examples of Faraday dispersion functions for a \textit{single baseline}. \textit{Top:} Faraday components only. \textit{Bottom:} The same Faraday components (F) with noise (N), leakage (L) and DC offsets (DC) added.}
\label{singlebeg}
\end{center}
\end{figure}

Assuming equal weights for all Stokes visibilities gives the following Mueller operator for the full interferometric system,
\begin{eqnarray}
\textbf{M} &=& \frac{1}{N(N-1)}\sum_i^N \sum_{j\neq i}^N \textbf{M}_{ij} \nonumber\\
&=& \frac{1}{N(N-1)}\sum_i^N \sum_{j\neq i}^N \textbf{S}(\textbf{J}_i \otimes \textbf{J}_j^*)\textbf{S}^{-1},
\end{eqnarray}
where both the baseline and its conjugate baseline are summed over in order to give a real-valued result. Using Equation~(\ref{eq:JStokes}), the Mueller matrix describing leakage for a \textit{single} baseline with parallel linear feeds (i.e., cross-dipole feeds in antenna $i$ are approximately parallel to those in antenna $j$) to first order in leakage and gain error is given by
\begin{eqnarray}
\label{Mij}
\textbf{M}_{ij} =
\textbf{I}_4 +
\frac{1}{2}\left(\begin{array}{cccc}
\gamma_{++} & \gamma_{+-} & \delta_{+-} & -i\delta_{-+}\\
\gamma_{+-} & \gamma_{++} & \delta_{++} & -i\delta_{--}\\
\delta_{+-} & -\delta_{++} & \gamma_{++} & i\gamma_{--}\\
-i\delta_{-+} & i\delta_{--} & -i\gamma_{--} & \gamma_{++}
\end{array}\right),
\end{eqnarray}
where $\textbf{I}_4$ is the $4 \times 4$ identity matrix,
\begin{eqnarray}
\gamma_{++} &=& (\Delta g_{i,x} + \Delta g_{i,y}) + (\Delta g^*_{j,x} + \Delta g^*_{j,y}),\\
\gamma_{+-} &=& (\Delta g_{i,x} - \Delta g_{i,y}) + (\Delta g^*_{j,x} - \Delta g^*_{j,y}),\\
\gamma_{--} &=& (\Delta g_{i,x} - \Delta g_{i,y}) - (\Delta g^*_{j,x} - \Delta g^*_{j,y}),\\
\delta_{++} &=& (d_{i,xy} + d_{i,yx}) + (d^*_{j,xy} + d^*_{j,yx}),\\
\delta_{+-} &=& (d_{i,xy} - d_{i,yx}) + (d^*_{j,xy} - d^*_{j,yx}),\\
\delta_{-+} &=& (d_{i,xy} + d_{i,yx}) - (d^*_{j,xy} + d^*_{j,yx}),\\
\delta_{--} &=& (d_{i,xy} - d_{i,yx}) - (d^*_{j,xy} - d^*_{j,yx}),
\label{d--}
\end{eqnarray}
and uniform, unit ideal gains\footnote{For non-uniform and/or non-unit gains the relevant gain factors follow through the algebra, skewing the Mueller matrix. This can be incorporated through the original choice of basis, leaving Equations~(\ref{Mij})--(\ref{d--}) unchanged.} have been used for convenience, such that
\begin{eqnarray}
g_{i,p} = 1 + \Delta g_{i,p}.
\end{eqnarray}
The resulting signal (again, for a single baseline only) including leakage and gain errors (\textit{primed}) that originates from a general input signal (\textit{non-primed}) is
\begin{eqnarray}
I^{\prime}_{ij} &=& (1 + \frac{1}{2}\gamma_{++})I_{ij} + \frac{1}{2}\gamma_{+-}Q_{ij} + \frac{1}{2}\delta_{+-}U_{ij} - \frac{1}{2}i\delta_{-+}V_{ij},\nonumber\\
\label{I_prime_b}
&&\\
Q^{\prime}_{ij} &=& (1 + \frac{1}{2}\gamma_{++})Q_{ij} + \frac{1}{2}\gamma_{+-}I_{ij} + \frac{1}{2}\delta_{++}U_{ij} - \frac{1}{2}i\delta_{--}V_{ij},\nonumber\\
\label{Q_prime_b}
&&\\
U^{\prime}_{ij} &=& (1 + \frac{1}{2}\gamma_{++})U_{ij} + \frac{1}{2}\delta_{+-}I_{ij} - \frac{1}{2}\delta_{++}Q_{ij} + \frac{1}{2}i\gamma_{--}V_{ij}.\nonumber\\
\label{U_prime_b}
&&
\end{eqnarray}
Figure~\ref{singlebeg} shows simulated examples of the Faraday dispersion function, $F(\phi)$, for a single baseline, simply illustrating the effect of leakage between Stokes~$Q$ and $U$, noise and DC offsets. In these examples, gain errors and leakage from Stokes~$I$ have been ignored. The amount of leakage from Stokes~$Q$ into Stokes~$U$ and Stokes~$U$ into Stokes~$Q$ is parameterised by $L_{QU}$ and  $L_{UQ}$ respectively (see Appendix~\ref{Uniform gains}). The error coffecients in these example have been exaggerated for demonstration. The upper plot shows the three Faraday components only, at depths $\phi = -100$, 250 and 300~rad\,m$^{-2}$. The sidelobe structures that are visibly associated with each peak originate from the finite bandwidth of the correlator and are scaled copies of the rotation measure spread function, $R(\phi)$, introduced in Section~\ref{Faraday dispersion function}. The relative magnitude of each peak depends on the intrinsic brightness temperature of the linearly polarised radiation that originates from behind and between the Faraday screens. The absolute magnitude of the peaks are dependent on the size (and sign) of the leakage and gain errors for that particular baseline. The lower plot shows the same Faraday components (labelled `F'), however, noise (labelled `N'), leakage (labelled `L') and DC offsets (labelled `DC') have been included. Power in the Faraday dispersion function due to noise appears in all Faraday depth channels. Note that unlike the effective Stokes~$I$ Faraday dispersion function, $F(\phi)$ is not Hermitian and therefore $|F(-\phi)| \neq |F(\phi)|$.

When considering all baselines, we simulate both gain and leakage errors by sampling a zero-mean, Gaussian distribution. The standard variation of these distributions have been chosen to be 10~per~cent of the nominal, unit gain for each polarisation and for each antenna.

\subsection{Instrumental calibration}
\label{Instrumental calibration}

We assume that a real-time calibrating procedure will produce a full polarisation calibration solution, including fits from polarised calibrators and constraints of most of the degrees of freedom in the primary beam models using non-polarised calibrators \citep{mitchell2008,lonsdale2009}. However, these calibration solutions will be produced from real-time analysis, in image-space and on short time scales ($\sim 10$~s). It can be assumed these solutions will only be stable over some minutes and not the total integration time of hundreds of hours spread over months/years. In addition, calibration solutions are to be applied on a baseline by baseline basis in order to obtain good polarisation purity and to prevent calibration errors from introducing artifacts into the snapshots. The instrument polarisation we address in this work, however, are the leakages and errors that do make their way into the snapshot observations and therefore the fully integrated EoR data sets, that may still be significantly larger than the cosmic signal. Therefore, using a single calibration solution is valid only for these short snapshots, where the S/N is very low, thus hiding the polarised foreground contamination in a forest of peaks (in all Stokes parameters). We overcome this by using a single calibration solution which we assume to represent some linear combination of known calibration solutions from all snapshots.

\subsection{Instrumental noise}
\label{Instrumental noise}

Thermal noise is present in every visibility measurement for every frequency channel and will be instrument dependent. Section~\ref{App:Instrumental noise} discusses how we model thermal noise. Since we are applying cleaning procedures to fully integrated data (as opposed to short snapshots), the rms of the instrumental noise is generally much smaller than the contamination by polarised foregrounds considered in this work. As a result, the fitting of peaks in Faraday dispersion functions is only slightly affected. We have found this to be the case when instrumental noise has been included in cleaning simulations. For this reason, we do not include thermal noise in our simulations, but rather, provide an instrumental sensitivity curve when considering the statistical detectability of the EoR signal (see Section~\ref{Spherically averaged power spectra}).

\section{Synthetic data cube}
\label{Synthetic data cube}

Having introduced the astrophysical sources and instrumental effects we include in our simulations, we now summarise the method with which these are combined in order to produce a full three-dimensional synthetic data set. This combination is performed in the Stokes representation and in image-space, however, some components, such as the cosmic signal, have been generated or modified or both in visibility-space (i.e., in the $uv$-plane of each frequency channel). The observation and instrumental parameters are initially specified. These include the bandwidth, $B$, central frequency, $\nu_0$, number of frequency channels, $N_{\rm ch}$, specific $uv$-tapering used and the parameters describing the Faraday screens.

We begin by simulating and combining the astrophysical sources, starting with the cosmic signal, in terms of brightness temperature in a cube that corresponds to the specified observation. The EoR signal appears in Stokes~$I$ only and is simulated using the method described in Section~\ref{Epoch of reionisation signal}. The simulation snapshot redshift, $z$, and comoving side length, $\Delta D$, of the EoR cube are given by
\begin{eqnarray}
z = \frac{\nu_{21}}{\nu_0} - 1
\end{eqnarray}
and
\begin{eqnarray}
\Delta D \approx \frac{c(1+z)^2B}{\nu_{21} H(z)},
\end{eqnarray}
where $\nu_{21}$ is the rest-frame frequency of the 21-cm line and $H(z)$ is the Hubble constant calculated at redshift $z$. The EoR simulation is then instrumentally smoothed in the sky-plane for each frequency channel using the method described in Section~\ref{The synthesised beam and beam depolarisation}.

Next, the total intensity of DGSE is simulated as per Section~\ref{Non-polarised DGSE}, and its signal is added to Stokes~$I$. Because we are assuming that there is no circularly polarised signal present, we set Stokes~$V$ to zero throughout.

The first effect to be simulated is that of thin Faraday screens. The corresponding polarised components of the DGSE are determined as per Section~\ref{Polarised DGSE}, the plane of polaristion is Faraday rotated and added to Stokes~$Q$ and $U$.  For a single Faraday screen the intrinsic degree of polarisation used in Equations~(\ref{Q_F}) and (\ref{U_F}) is simply the total brightness temperature of the polarised DGSE, i.e., $f T_{\rm b}^{\rm G}(\nu)$. This corresponds to the scenario where all of the polarised signal originates from behind the Faraday screen. In the scenario of multiple Faraday screens, the sum of intrinsic degrees of polarisation for each screen equal the total brightness temperature of the polarised DGSE. \cite{brentjens2005} provide an excellent illustration of more complicated emission and Faraday rotating scenarios. As previously mentioned, for simplicity and to illustrate the principle behind our cleaning method, we simulate the effect of a single thin Faraday screen only.

Finally, the effect of instrumental polarisation leakage is simulated, using instrumental errors of the form discussed in Section~\ref{leakage}, for each baseline.

\section{Foreground subtraction}
\label{Foreground subtraction}

In this section, we discuss the methods of removing diffuse Galactic foregrounds. By including only diffuse Galactic synchrotron emission we assume that all resolved discrete and extended sources have already been successfully removed from the observed maps, without any subtraction residuals. The requirements for subtracting point sources to a sufficient level are discussed by \cite{liu2009}, and a method for achieving this is presented by Pindor et al. \citep{pindor2010}.

\subsection{Continuum foreground subtraction}
\label{Continuum component subtraction}

There has been significant discussion in the literature regarding low-frequency continuum foregrounds and strategies for their removal. While fluctuations due to these foregrounds are expected to be several orders of magnitude larger than the reionisation signal \citep[e.g.,][]{dimatteo2002,oh2003}, the spectra of these foregrounds are anticipated to be smooth. As the reionisation signature includes fluctuations in frequency as well as angle, it has therefore been proposed that these foregrounds be removed through continuum subtraction \citep[e.g.,][]{gnedin2004,wang2006,jelic2008}, or using the differences in symmetry from power spectra analysis \citep[][]{morales2004,zaldarriaga2004}.

We model the foreground continuum using a polynomial of the form
\begin{eqnarray}
\log_{10} (T_{\rm b}^{\rm G}) = \sum_{i=0}^n c_i [\log_{10}(\nu)]^i,
\end{eqnarray}
where $\nu$ is the observed frequency and $n$ is the highest order of $\log_{10}(\nu)$ used for fitting. The appropriateness of the functional form of this fit can be illustrated by considering a pure DGSE signal along a single line of sight, with no instrumental polarisation leakage or noise. In this case, we can perfectly fit our foreground model using a quadratic ($n = 2$) in $\log_{10}(\nu)$ with
\begin{eqnarray}
\label{fit_solutions}
c_0 &=& \log_{10}(C) + \alpha_{\rm syn}\log_{10}(\nu_0) - \Delta\alpha_{\rm syn}[\log_{10}(\nu_0)]^2,\\
c_1 &=& -\alpha_{\rm syn} + 2\Delta\alpha_{\rm syn}\log_{10}(\nu_0),\\
c_2 &=& -\Delta\alpha_{\rm syn},
\end{eqnarray}
where $C = \bar{T}^{\rm G}_{\rm b,0} + \delta T^{\rm G}_{\rm b,0}(\vec{\theta})$.

In the general case, we perform fits with this functional form along the line of sight for each spatial pixel in the simulated sky-plane. We then subtract the best fit, leaving residual fluctuations around the foreground emission with a near zero-mean brightness temperature, which will include any residual continuum foregrounds, leaked polarised signal, instrumental noise and the EoR signal. The consequence of removing the foreground continuum, together with the fact that interferometers do not make zero-spacing measurements, means that the true mean signal from each line of sight is unknown. As will be shown, this results in a decreased contrast between ionised and non-ionised regions, and a loss of power from large-scale (small wavenumber) modes \citep{geil2008b}.

\subsection{Polarised foreground subtraction}
\label{Polarised component subtraction}

\begin{figure}
\includegraphics[width=8.5cm]{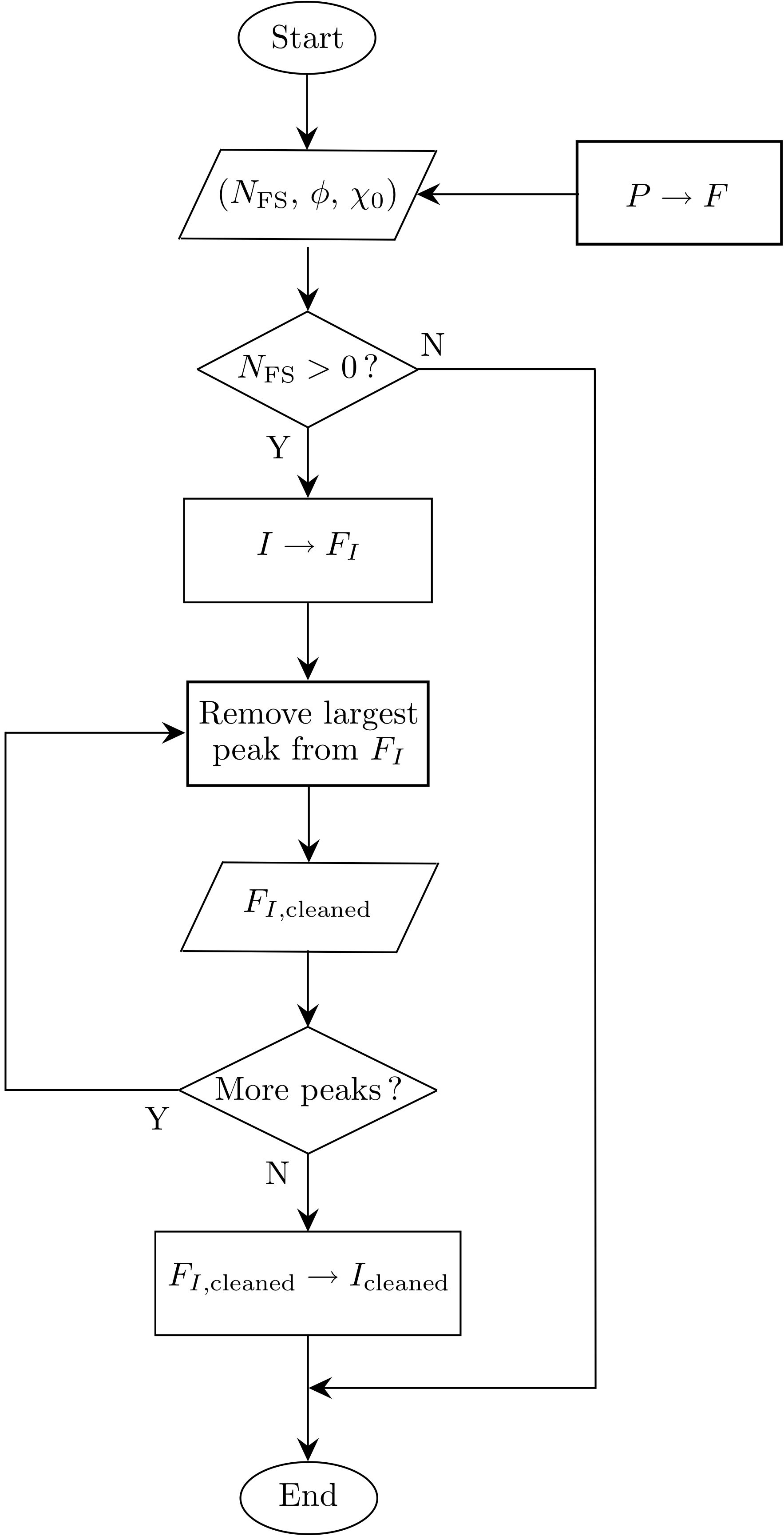} 
\caption{Flow diagram for single line-of-sight polarised foreground removal.}
\label{flow}
\end{figure}

As discussed in Section~\ref{leakage}, the leakage of polarised foreground components into the Stokes~$I$ signal can lead to significant contamination of the cosmic signal. The procedure for removing this contamination is shown schematically in Figure~\ref{flow}. For each line of sight:
\begin{enumerate}
\item the observed complex linear polarisation, $P(\lambda^2)$, is Fourier transformed to find its corresponding Faraday dispersion function, $F(\phi)$;
\item the number of peaks, $N_{\rm FS}$, is determined from $F(\phi)$ by a thresholding technique or a peak removal procedure similar to the method discussed here ($F$ is used rather than $F_I$ since the signal-to-noise ratio is greater and the polarised signal is `pure' in $F$), and each peak's Faraday depth and intrinsic polarisation angle is found using Equation~(\ref{eq:p and chi});
\item in the case that $N_{\rm FS} = 0$, there is no polarised contamination to clean; otherwise,
\item we attempt to remove each identified peak from the Stokes~$I$ signal. This is done by Fourier transforming $I(\lambda^2)$ to give the effective Stokes~$I$ Faraday dispersion function, $F_I(\phi)$, then removing the largest peak by constructing a model of the real and imaginary parts of a polarised component (using the $\phi$ and $\chi_0$ found from the analysis of $F$ and leakage solutions found by calibration);
\item the cleaned $F_I(\phi)$ is inverse Fourier transformed back to give the cleaned Stokes~$I$ signal.
\end{enumerate}
This process is repeated for each line of sight, and the cleaned Stokes~$I$ signals are reconfigured into a three-dimensional data cube. It is possible to remove polarised foreground components in a number of iterations using a loop-gain factor $\gamma$ to bring the peaks down slowly, treating the strongest component in each iteration. This method is similar in spirit to the \texttt{CLEAN} deconvolution algorithm \citep{hogbom1974} used in radio imaging. However, we subtract a peak in a single iteration in our simulations. 

The single iteration procedure described above performs well for lines of sight where the number of Faraday screens is correctly calculated. This is the case when the RM peaks are spaced sufficiently far apart in $\phi$-space such that one does not superimpose another (the differentiation between peaks also depends on the rotation measure spread function, $R(\phi)$, of the instrument at a particular frequency). The single iteration procedure fails when the different Faraday screens are close to each other, however, follow-up analysis of lines of sight, including higher dimensional fitting, where this is an issue can be performed.

\subsection{Summary of assumptions and other considerations}
\label{Summary of assumptions}

The aim of this paper is to present the important first step toward solving the important problem of polarised foreground leakage. In order to do this a number of assumptions have been made throughout this work so as to simplify the simulations and attempt to demonstrate the basic principal of our algorithm.
\begin{enumerate}
\item
We have only simulated a single thin Faraday screen as a means to simply demonstrate our idea of leakage removal.  The real situation is of course more complicated, including multiple thin screens and/or slabs, but we hope to tackle a more realistic scenario in later work.
\item
The effect of the instrumental beam has been neglected. Any effect of the beam should of course be well understood and modelled. We assume that these effects have been treated prior to the application of our procedure.
\item
There are many contaminants and foregrounds to consider, however, any components not mentioned already have been considered insignificant to the present procedure.
\item
We have not made a realisation of a power-spectrum for the spatial fluctuations in non-polarised foreground. A Gaussian field has been used for the sake of simplicity.
\item
Observations in our simulations are assumed to have been made in the zenith direction. In a more realistic model observations away from the zenith would need to be considered and the impact of dipole projection considered.
\item
The size of the errors we consider is no doubt larger than what be actually be the case. This was chosen in order to make a clear demonstration of the leakage. Also, we have modelled the errors with no bias despite the possibility that in reality a systematic bias may be present due to the mass-manufacturing method of the dipoles.
\item
We make an important assumption that Faraday rotation effects of the ionosphere have already been accounted for. We argue that these ionosphere through which the instrument is observing is well measured and understood through the use of polarised point-source callibrators and that any effect has been removed (by our method or otherwise) prior to the application of our procedure on Galactic Faraday rotating regions.
\end{enumerate}

Naturally, at some point when the level of noise is similar to the level of polarised foreground leakage in the Faraday dispersion function a confusion limit will be found. This has been tackled in the following way. First, by fitting for RM and polarisation angle using $F$ (rather than $F_I$) where the level of confusion will be less since the peaks corresponding to Faraday-rotating structures will be larger than their Stokes~$I$ counterparts. Secondly, since the data cubes comprise of integrated data, the level of noise can, to a certain degree, be controlled by the integration time.

\section{Spherically averaged power spectra}
\label{Spherically averaged power spectra}

In this paper, we employ the dimensionless power spectrum, $\Delta_{21}^2(k,z)$, as the key statistical measure, which is the contribution to the variance of the redshifted 21-cm brightness temperature contrast, $\delta T_{\rm b}(z)$, per logarithmic interval in wavenumber, $k = 2\pi/\lambda$. This measure is related to the dimensional form of the power spectrum, $P_{21}(k,z)$, by
\begin{eqnarray}
\label{Deltasq}
\Delta_{21}^2(k,z) = \frac{1}{(2\pi)^3} 4\pi k^3P_{21}(k,z).
\end{eqnarray}
$P_{21}(k,z)$ is estimated by averaging over all $m$ modes of the Fourier transform ($\hat{T}$) of $T(z)$ in a thin spherical shell in $k$-space,
\begin{eqnarray}
P_{21}(k,z) \equiv \langle|\hat{T}(\kvec,z)|^2\rangle_k = \frac{1}{m}\sum _{i=1}^m |\hat{T}_i(k,z)|^2.
\end{eqnarray}
Note that we must use caution when employing this statistical measure since the spherical symmetry of the signal is broken by redshift evolution over certain ranges of redshift ($\Delta z \approx 2$ for a 32~MHz band centred on $z = 8.5$). This issue has been discussed by a number of authors \citep[see, e.g.,][]{morales2004,bl2005,mcquinn2006} and its effect on the sensitivity of the 21-cm power spectrum has been investigated by \cite{mcquinn2006}. Following the removal of both the continuum and polarised foregrounds, the full 32~MHz data cube is divided into four 8~MHz sub-bands ($\Delta z \lsim\,\,0.5$ for $z_{\rm ov} < z < 8.5$), each of which is independently reduced into a power spectrum. This sub-banding, together with continuum foreground removal, eliminates measurements of the largest spatial scales along the line-of-sight direction (these corresponding to the lowest $k_{\parallel}$ modes) and imposes a minimum accessible wavenumber of $k_{\rm min} \approx 0.04[(1+z)/7.5]^{-1}$~Mpc$^{-1}$].

\subsection{Sensitivity to the 21-cm power spectrum}
\label{Sensitivity to the 21-cm power spectrum}

Our numerical power spectrum measurements are subject to sample variance, arising from the finite number of independent modes counted in each $k$-shell, which corresponds to the finite number of independent wavelengths that can fit into the simulated volume. Any radio interferometer is also subject to instrumental noise and has limited sensitivity arising from the finite volume of the observation. We consider observational parameters corresponding to the design specications of the MWA, and of a hypothetical follow-up to the MWA (which we term the MWA5000). The MWA 5000 is assumed to follow the basic design of the MWA. The quantitative differences are that we assume the MWA5000 to have 5000 tiles within a diameter of 2~km, with a flat antenna density core of radius 40~m. In each case, we assume one field is observed for an integrated time of 1000~hr.

We compute the sensitivity with which the 21-cm power spectum could be detected following the procedure outlined by \cite{mcquinn2006} and \cite{bowman2006}. Section~\ref{App:Power spectrum measurement} of the Appendix contains a full derivation of the sensitivity-related quantities used. Written in terms of the cosmic wavevector $\kvec = \kvec_\parallel + \kvec_\perp$, where $\kvec_\parallel$ and $k_\perp$ are the components of $\kvec$ parallel and perpendicular to the line of sight respectively (see Appendix \ref{App:Power spectrum measurement}), the resulting error in the 21-cm power spectrum \textit{per mode} is
\begin{eqnarray}
\delta P_{21}(\kvec) = \frac{T_{\rm sys}^2}{Bt_0} \frac{D^2 \Delta D}{n_{\rm b}(U,\nu)} \left(\frac{\lambda^2}{A_{\rm e}}\right)^2  + P_{21}(\kvec),
\end{eqnarray}
where $T_{\rm sys} \approx 250[(1+z)/7]^{2.6}$~K is the system temperature of the instrument, $D(z)$ is the comoving distance to the point of emission at redshift $z$, $\Delta D$ is the comoving depth of the survey volume corresponding to the bandwidth $B$, $t_0$ is the total integration time, $n_{\rm b}(U,\nu)$ is the number density of baselines that can observe the visibility $\Uvec$, where $U = k_\perp D/2\pi$ and $\lambda$ is the observed wavelength.

Although the observed 21-cm power spectrum is not spherically symmetric, it is symmetric about the line of sight. This makes it possible to calculate the overall power-spectral sensitivity of the radio interferometer using the Fourier modes contained within an infinitesimal annulus around the line of sight of constant $(k,\theta)$, where $\cos(\theta) = \kvec \cdot \hat{\zvec}/k$ ($\hat{\zvec}$ is the unit vector pointing in the direction of the line of sight). The power spectral sensitivity over such an annulus is given by
\begin{eqnarray}
\label{sigmaP}
\sigma_P(k,\theta) = \frac{\delta P_{21}(k,\theta)}{\sqrt{N_{\rm m}(k,\theta)}},
\end{eqnarray}
where $N_{\rm m}(k,\theta)$ denotes the number of observable modes in the annulus (only modes whose line-of-sight components fit within the observed bandpass are included). In terms of the $k$-vector components $k$ and $\theta$, the number of independent Fourier modes within an annulus of radial width $dk$ and angular width $d\theta$ is $N_{\rm m}(k,\theta) = 2\pi k^2 V \sin(\theta) dk\,d\theta/(2\pi)^3$, where $V = D^2\Delta D(\lambda^2/A_{\rm e})$ is the observed volume. Averaging $\sigma_P(k,\theta)$ over $\theta$ gives the spherically averaged sensitivity to the 21-cm power spectrum $\sigma_P(k)$,
\begin{eqnarray}
\frac{1}{\sigma^2_P(k)} = \sum_{\theta}\frac{1}{\sigma^2_P(k,\theta)}.
\end{eqnarray}
In order to find the sensitivity in terms of $\Delta^2_{21}$ we use Equation~(\ref{Deltasq}), which gives
\begin{eqnarray}
\sigma_{\Delta^2}(k,z) = \frac{1}{(2\pi)^3}4\pi k^3 \left[\sum_{\theta}\frac{1}{\sigma^2_P(k,\theta,z)}\right]^{-1/2}.
\end{eqnarray}

\section{Results}
\label{Results}

In this section, we present the results of our simulations and analysis. Section~\ref{Single line of sight} shows the result of applying the polarised foreground removal algorithm for a single line of sight. Section~\ref{Full data cube} gives the results for a full three-dimension data set. Figure~\ref{Spherically averaged power spectra} shows the resulting spherically averaged 21-cm power spectra.

\subsection{Single line of sight}
\label{Single line of sight}

\begin{figure*}
\begin{center}
\includegraphics[width=17cm]{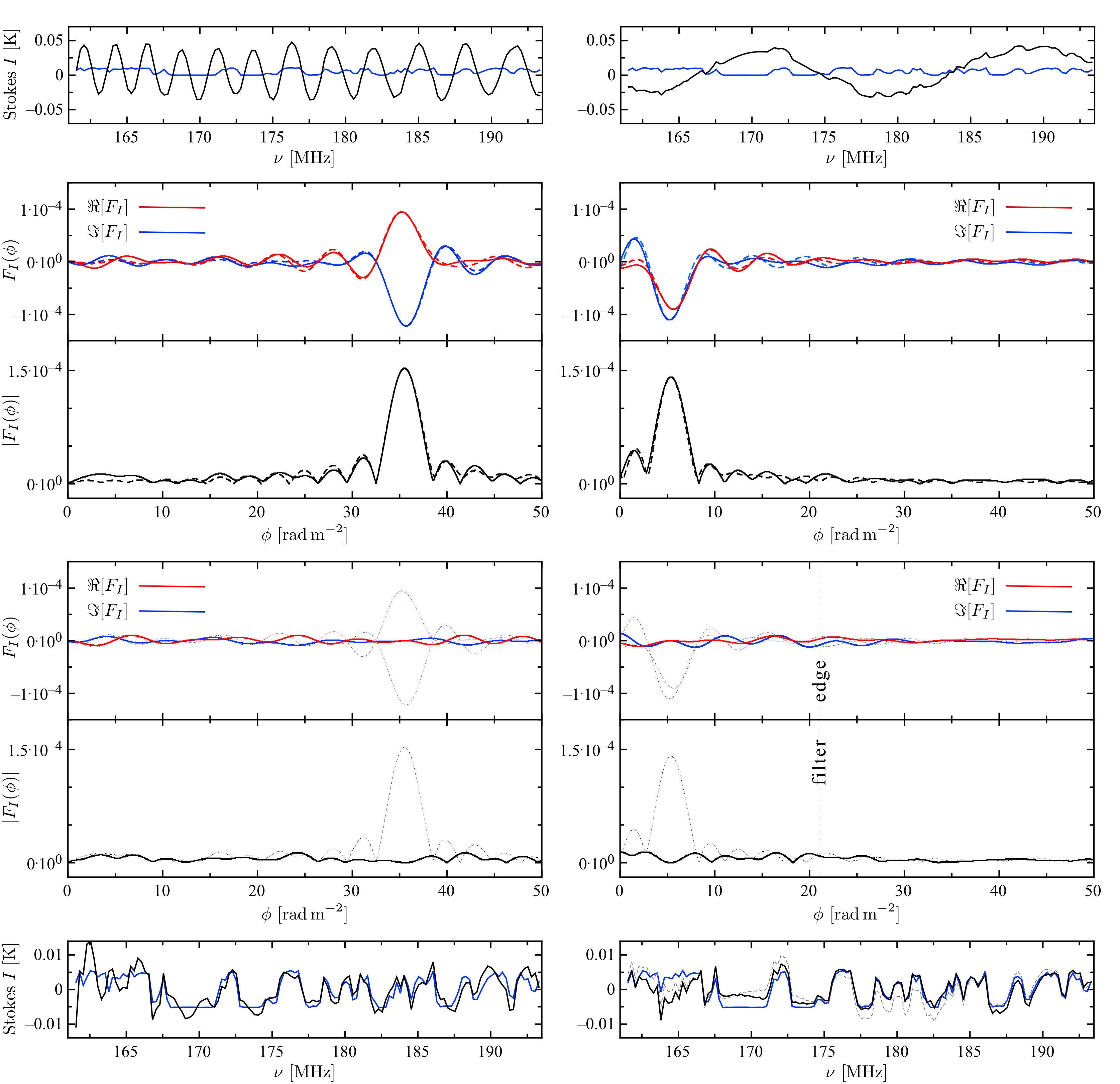} 
\caption{Examples of the removal of a single polarised foreground component from Stokes~$I$. \textit{Top:} Synthetic Stokes~$I$ signal (\textit{black}) comprising of EoR signal (\textit{blue}) and contamination due to instrumental polarisation leakage. \textit{Second from top:} Effective Stokes~$I$ Faraday dispersion function (\textit{solid}) of total signal and fits to its magnitude and real and imaginary parts. (\textit{dashed}). \textit{Second from bottom:} Residual effective Stokes~$I$ Faraday dispersion function after cleaning. Original effective Stokes~$I$ Faraday dispersion function (\textit{faint}) is shown for comparison. \textit{Bottom:} Cleaned Stokes~$I$ signal (\textit{black}) and EoR signal with line-of-sight mean removed (\textit{blue}) for comparison. \textit{Left:} Single Faraday screen at $\phi = 35.5$~rad\,m$^{-2}$ with $p(\lambda^2_0) \approx 176.8$~K and $\chi_0 \approx 0.794$. \textit{Right:} Single Faraday screen at $\phi = 5.5$~rad\,m$^{-2}$ with $p(\lambda^2_0) \approx 161.6$~K and $\chi_0 \approx -1.52$. A high-pass filter has been applied to the $\phi = 5.5$~rad\,m$^{-2}$ example in order to eliminate the residual long-`wavelength' fluctuation that remains after cleaning (\textit{faint dashed}). The filter edge at $\phi_{\rm f} \approx 21.2$~rad\,m$^{-2}$ is shown by the vertical line on the plot of the cleaned effective Stokes~$I$ Faraday dispersion function.}
\label{clean_1d_eg}
\end{center}
\end{figure*}

Figure~\ref{clean_1d_eg} demonstrates the removal of a single polarised foreground component from a noiseless Stokes~$I$ signal. For this example, we assume that any continuum DGSE has already been removed, leaving only the cosmic signal and leaked polarised foreground. Two different values of Faraday depth have been modelled in order to show the difference in cleaning performance. The left-hand example uses $\phi = 35.5$~rad\,m$^{-2}$ with $p(\lambda^2_0) \approx 176.8$~K and $\chi_0 \approx 0.794$. The right-hand example uses $\phi = 5.5$~rad\,m$^{-2}$ with $p(\lambda^2_0) \approx 161.6$~K and $\chi_0 \approx -1.52$.

The top Stokes~$I$ visibility versus frequency plots in Figure~\ref{clean_1d_eg} show the signal to be cleaned (\textit{black}) and the model cosmic signal (\textit{blue}) for reference. Note that the frequency axis is linear and therefore the apparent `period' of the contaminated signal containing the leaked polarised foreground is non-uniform. This is because of the $\lambda^2$-dependence of the plane of polarisation as per Equation~(\ref{eq:chi}). The second row of plots show the effective Stokes~$I$ Faraday dispersion function, $F_I(\phi)$, of the contaminated signal, including the real and imaginary part as well as its modulus. The fit to each of these components (\textit{dashed}) are shown for comparison. The third row of plots show the cleaned effective Stokes~$I$ Faraday dispersion function with the contaminated effective Stokes~$I$ Faraday dispersion function (\textit{faint}) shown for comparison. The bottom row of plots show the cleaned signal (\textit{black}) after being Fourier inverted and the model cosmic signal (\textit{blue}) for reference.

It is evident that the instrumental polaristion leakage due to the Faraday screen at the smaller depth (\textit{shallow} Faraday screen) is less effectively removed than the deeper screen. This is due in part to a residual long-`wavelength' fluctuation in the signal after continuum foreground removal which leaves small-$\phi$ power in the effective Stokes~$I$ Faraday dispersion function. Residuals such as these can make a small-$\phi$ peak in the Stokes~$I$ Faraday dispersion function asymmetric and therefore more difficult to fit accurately. Deeper peaks are not affected in this way as the modelled continuum foreground does not contain fluctuations that masquerade at these Faraday depths, therefore no residuals at these depths are present after continuum subtraction making more accurate fitting possible.

Unwanted small-$\phi$ residuals have been removed by applying a high-pass filter to the cleaned effective Stokes~$I$ Faraday dispersion function. The extent of this filter is $0 \leq \phi_{\rm f} \lsim\,\,21.2$~rad\,m$^{-2}$, which has an equivalent width in frequency-space of approximately 4.7~MHz for the correlator specifications used here. Negligible cosmic signal is removed by this procedure since the EoR fluctuations are of shorter `wavelength' at the corresponding redshift at this frequency.

\subsection{Full data cube}
\label{Full data cube}

\begin{figure*}
\includegraphics[width=17cm]{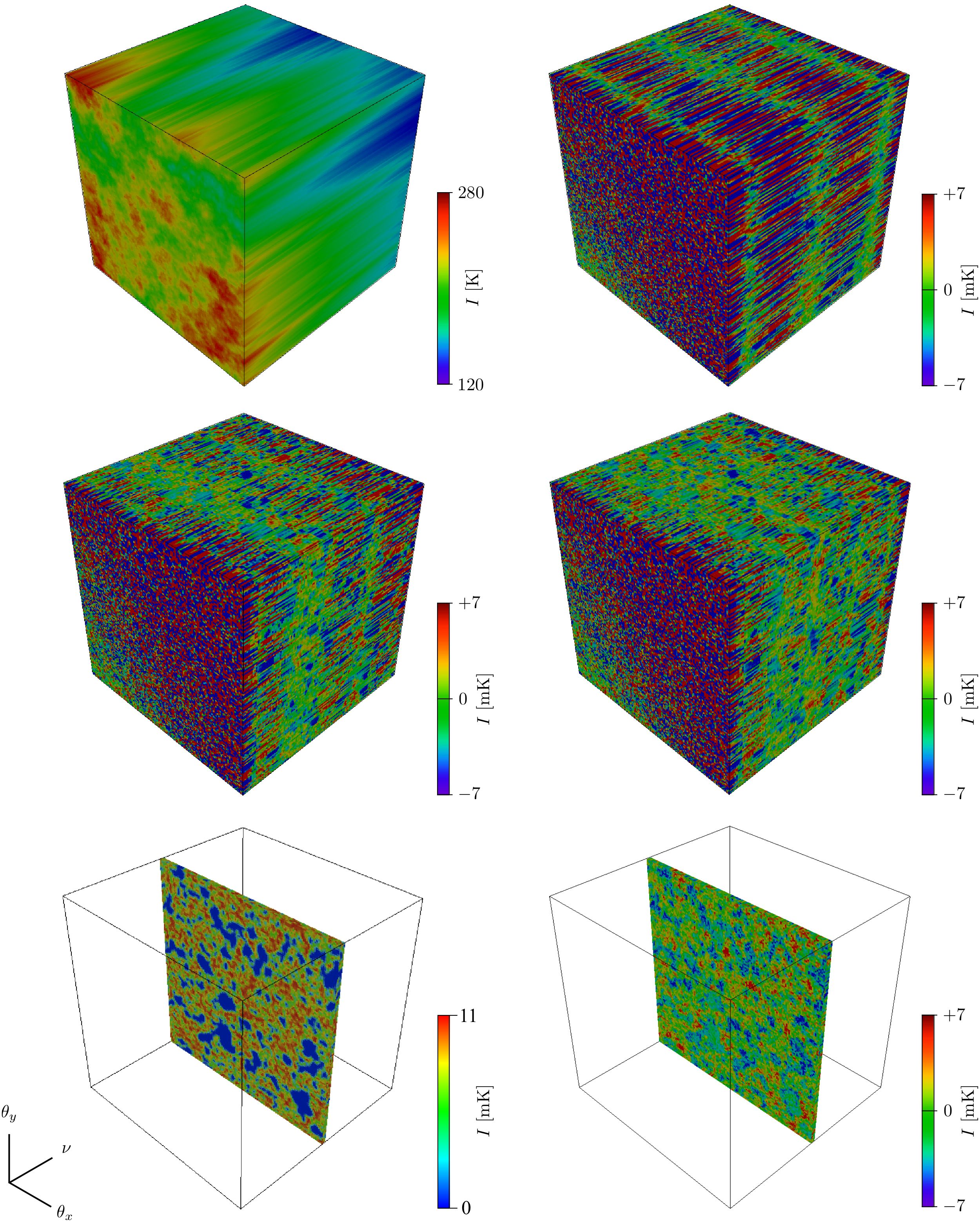} 
\caption{Simulated noiseless Stokes~$I$ signal over a 32~MHz band, centred on 177.5~MHz ($6.3 \lsim\,\,z \lsim\,\,7.8$) using 128 channels. Faraday depth has been modelled as a gradient over sky-plane with $1 \leq \phi \leq 5$~rad\,m$^{-2}$. Intrinsic polarisation angle for each line of sight has been randomly assigned. \textit{Top-left:} Full noiseless Stokes~$I$ signal showing domination by DGSE. \textit{Top-right:} Stokes~$I$ signal after continuum foreground removal. \textit{Middle-left:} Stokes~$I$ signal after continuum and polarised foreground removal. \textit{Middle-right:} Stokes~$I$ signal after continuum and polarised foreground removal and high-pass filtering. \textit{Bottom-left:} Model EoR signal. \textit{Bottom-right:} Central channel slice through Stokes~$I$ signal after continuum and polarised foreground removal and high-pass filtering.}
\label{plots_z7-0}
\end{figure*}

Figure~\ref{plots_z7-0} shows a simulated three-dimensional Stokes~$I$ data cube over a 32~MHz band centred on 157.8~MHz ($6.3 \lsim\,\,z \lsim\,\,7.8$) using 128 channels. This simulation includes both continuum and polarised foreground removal.  Faraday depth has been modelled as a gradient over sky-plane with $1 \leq \phi \leq 5$~rad\,m$^{-2}$ and the intrinsic polarisation angle for each line of sight has been randomly assigned, $-\pi/2 < \chi \leq \pi/2$~rad. The upper-left cube shows the full Stokes~$I$ signal before cleaning. The upper-right cube has had the continuum DGSE fitted and removed in each line of sight, leaving the cosmic signal and the effects of instrumental polarisation leakage. The banding along the $\theta_x$- and $\theta_y$-frequency planes result from the oscillatory behaviour of the leaked polarised signal. The middle-left cube is the raw fully cleaned signal having had the leaked polarised signal removed. The pre-leakage-cleaned banding has mostly disappeared, with the exception of channels close to the edge of the frequency band. This is a systematic effect due in part to the continuum foreground removal. The central-right cube shows the fully cleaned signal after high-pass filtering of the effective Stokes~$I$ Faraday dispersion function of each line of sight. The same filter width has been used as for the one-dimensional example in Section~\ref{Single line of sight}. The central frequency slice of the fully cleaned and high-pass filtered cube is shown at the lower-right, with the model EoR signal shown at the lower-left for comparison. The model EoR cube has not had each line of sight's mean subtracted, which would occur as a consequence of the continuum foreground subtraction.

\begin{figure*}
\includegraphics[width=10cm]{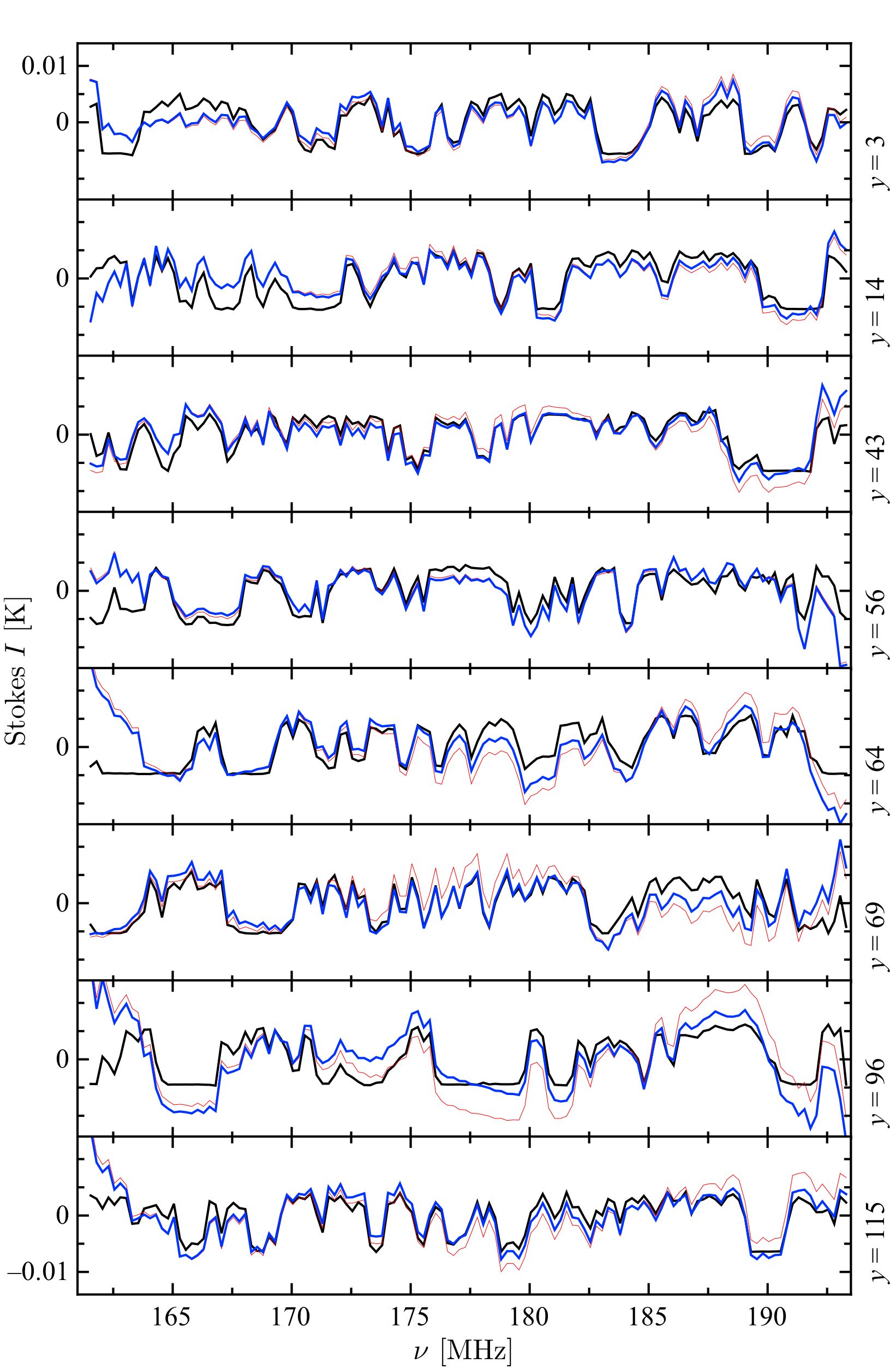} 
\caption{Simulated noiseless Stokes~$I$ signal over a 32~MHz band, centred on 177.5~MHz ($6.3 \lsim\,\,z \lsim\,\,7.8$) using 128 channels. Faraday depth has been modelled as a gradient over sky-plane with $1 \leq \phi \leq 5$~rad\,m$^{-2}$. Intrinsic polarisation angle for each line of sight has been randomly assigned. Example line-of-sight signal profiles: model EoR signal with line-of-sight means removed (\textit{black}), fully cleaned signal with no high-pass filtering of their effective Stokes~$I$ Faraday dispersion function (\textit{red}) and fully cleaned signal after high-pass filtering of their effective Stokes~$I$ Faraday dispersion function (\textit{blue}).}
\label{plots_z7-2}
\end{figure*}

Figure~\ref{plots_z7-2} shows a selection of line-of-sight signal profiles from the three-dimensional simulation for the model cosmic signal (\textit{black}), fully cleaned signal with no high-pass filtering of their effective Stokes~$I$ Faraday dispersion function (\textit{red}) and fully cleaned signal after high-pass filtering of their effective Stokes~$I$ Faraday dispersion function (\textit{blue}).

\subsection{Power spectrum measurement in the presence of foregrounds}
\label{Power spectrum measurement in the presence of foregrounds}

\begin{figure*}
\includegraphics[width=9cm]{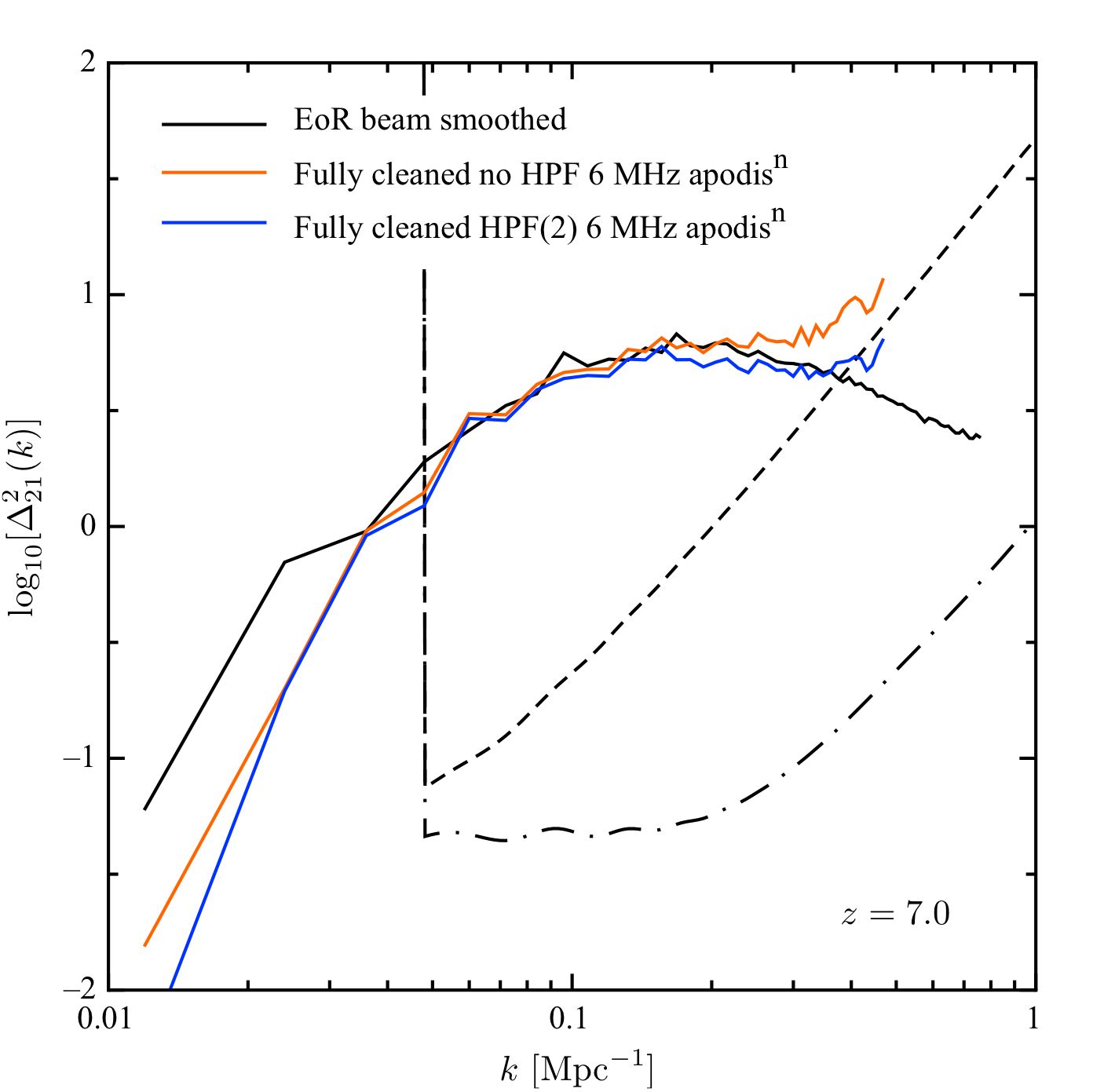} 
\caption{Spherically averaged 21-cm power spectra. Beam-smoothed model EoR signal; Stokes~$I$ signal after continuum and polarised foreground removal, high-pass filtering and band apodisation at redishifts $z = 7$. Instrumental sensitivity curves for MWA500 (\textit{dashed}) and MWA5000 (\textit{dot-dashed}) and 1000~hr integrations are also shown, where it has been assumed that the total 32~MHz bandwidth has been split into four 8~MHz sub-bands.}
\label{fig:ps}
\end{figure*}

Here, we summarise the effects of various foreground cleaning methods on the potential power spectrum measurement. Figure~\ref{fig:ps} shows the spherically averaged 21-cm power spectra for a foreground cleaning run performed on a simulated, noiseless Stokes~$I$ signal over a 32~MHz band, centred on 177.5~MHz ($6.3 \lsim\,\,z \lsim\,\,7.8$) using 128 channels. Instrumental sensitivity curves for MWA500 (\textit{dashed}) and MWA5000 (\textit{dot-dashed}) and 1000~hr integrations are also shown, where it has been assumed that the total 32~MHz bandwidth has been split into four 8~MHz sub-bands.

It can be seen that the power spectrum of the cleaned signal agrees very well with the power spectrum of the model cosmic signal, once it has been apodised and further cleaned with a high-pass filter, within the sensitivity limits of the MWA. The difference between the two for $k \lsim 0.04$~Mpc$^{-1}$ is due to the loss of large-scale contrast owing to continuum foreground subtraction. The up-turn at large $k$ (small scales) arises from the residual fluctuations present at the edges of the bandpass (as a result of polarised foreground removal).

Apodisation of the bandwidth is required in order to eliminate spurious power in the power spectra due to the band-edge effects discussed in Section~\ref{Full data cube}. No overall power is lost through this procedure, however, because the sampled volume is smaller, there will be a decreased effective sensitivity due to decreased number of Fourier modes, $N_{\rm m} \propto V$ [see Equation~(\ref{sigmaP})].

\section{Summary}
\label{Summary}

In this paper, we have shown how polarised signals can contaminate the non-polarised signal---and therefore contaminate the signature of cosmic reionisation---through the process of instrumental polarisation leakage. We have also demonstrated how, in principle, RM synthesis may be used to recover the cosmic signal by applying an RM synthesis technique to a three-dimensional synthetic data cube. In demonstrating this, we have assumed that the brightest point sources have been sufficiently removed, and modelled the brightest astrophysical sources of foreground contamination only, for the simplest case of a single, thin Faraday screen.

Our results show that it is possible to accurately recover the signature of reionisation late in the epoch of reionisation ($z \approx 7$), by way of the 21-cm power spectrum, within the sensitivity limits of an instrument similar to the MWA. In addition to the statistical signature of cosmic reionisation, we have also shown that the tomographic image of the percolating IGM can be successfully cleaned of contamination.

We have applied our technique using a simple foreground model and a single thin Faraday screen. We believe it is possible to extend the application of this technique to more complex and realistic foreground and Faraday screen models by performing the cleaning process iteratively. This is an important improvement for lines of sight that have multiple Faraday-rotating screens or screens that are thick enough to be resolved by the instrument. For situations where a line of sight contains multiple screens, well separated in $\phi$-space, the single iteration cleaning method used in this paper may be applied to each foreground peak in the effective Stokes~$I$ Faraday dispersion function. Although contamination of the Stokes~$I$ signal due to very shallow Faraday screens (which have fluctuations in frequency-space on a scale larger than the EoR signal) can be removed by simply applying a high-pass filter in $\phi$-space with minimal effect on the recovered cosmic signal, Faraday screens that lead to contamination that fluctuate on scales similar to the EoR signal (large $|\phi|$) cannot, but can be easily removed using our method.

It has already been shown that neither point sources, continuum DGSE nor instrumental noise will pose insurmountable obstacles to the detection of the cosmic 21-cm signal \citep[see, e.g.,][]{wang2006,geil2008b,jelic2008,bowman2009,liu2009}. Our results suggest that the cleaning of leaked polarised foregrounds will also be possible. Removing this contaminant will be a critical element in the removal of foregrounds that will be essential in order to detect the cosmological 21-cm signal.
 
\bigskip

\noindent {\bf Acknowledgments}
The authors would like to thank the anonymous referee for providing useful and detailed suggestions which has improved the manuscript. PMG acknowledges the support of an Australian Postgraduate Award and the hospitality of the University of Sydney, where part of this research was done. JSBW acknowledges the support of the Australian Research Council. BMG acknowledges the support of a Federation Fellowship from the Australian Research Council through grant FF0561298. The Centre for All-sky Astrophysics is an Australian Research Council Centre of Excellence, funded by grant CE11E0090.

\bibliography{bib.bib}

\appendix
\label{app:sensitivity}
\section{Statistical errors and instrumental sensitivity}
\label{Statistical errors and instrumental sensitivity}

\subsection{Fiducial array configuration}
\label{Fiducial array configuration}

\subsubsection{Coordinates and notation}
\label{Coordinates and notation}

\noindent The visibility measurements of the next generation of low-frequency radio arrays will sample the instrumental response from an inherently three-dimensional volume of space at high redshift. Only the fluctuating part of the signal contributes to an interferometric visibility, which we denote by $\Delta V(u,v,\nu)$, where $u$ and $v$ have their usual meaning \citep[see, e.g.,][]{tms}. As discussed by \cite{morales2005}, there are three useful representations of the data: the image representation, the visibility representation, and the full Fourier representation. The image cube is coordinated by sky position and frequency $(\theta_x,\theta_y,\nu)$ and can be transformed into a cube of visibilities by performing a two-dimensional Fourier transform in each frequency channel. Instrumental visibilities have coordinates given by $(u,v,\nu)$, where $u = 1/\theta_x$ and $v = 1/\theta_y$. The visibility cube is transformed into the full Fourier representation by performing a one-dimensional Fourier transform in frequency, giving coordinates in terms of $(u,v,\eta)$, where $\eta = 1/\nu$ is the wavenumber corresponding to the frequency of the visibility measurement. Here, we preserve the traditional meaning of the interferometric quantities $u$ and $v$ by defining the three-dimensional wavevector of a Fourier mode by
\begin{eqnarray}
\Upsilonvec = u\hat{\xvec} + v\hat{\yvec} + \eta\hat{\zvec} = \Uvec + \eta\hat{\zvec}.
\end{eqnarray}
The cosmological Fourier mode $\kvec = \kvec_\perp + \kvec_\parallel$ and its relationship to the instrumental Fourier mode $\Upsilonvec$ is discussed in Section~\ref{App:Power spectrum measurement}.

\subsubsection{Antenna distribution - continuous approximation}
\label{Antenna distribution - continuous approximation}

\noindent Although the number density of antennas for an interferometer will be discrete in practice, here we make the simplifying assumption of a continuous distribution. The continuous antenna distribution generalises our analysis (to some limited degree) and is justified for large numbers of antennas in a relatively compact configuration (i.e., a sensible distribution, considering the operating frequency, thereby providing full $uv$-coverage). For a circularly symmetric array of $N_{\rm a}$ antennas (tiles) with a constant antenna density core of radius $r_{\rm c}$, maximum radius $r_{\rm max}$ and an inverse square antenna density profile for $r>r_{\rm c}$, the properly normalised antenna number density as a function of $r$ is given by
\begin{eqnarray}
n_{\rm a}(r) = \left\{
\begin{array}{ll}
n_{\rm c} & r \leq r_{\rm c}\\
n_{\rm c}\left(\dfrac{r_{\rm c}}{r}\right)^2 & r_{\rm c} < r < r_{\rm max}\\
0 & r \geq r_{\rm max}
\end{array} \right.
\end{eqnarray}
where $n_{\rm c}$ is the antenna number density of the core, given by
\begin{eqnarray}
n_{\rm c} = \frac{N_{\rm a}}{\pi r_{\rm c}^2[1+2\ln(r_{\rm max}/r_{\rm c})]}.
\end{eqnarray}
This is capped such that there is a maximum density of one antenna per 36~m$^2$. The baseline number density can be written
\begin{eqnarray}
n_{\rm b}(U,\nu) = C_{\rm b}\int_{0}^{r_{\rm max}} 2\pi r dr\,n_{\rm a}(r) \int_{0}^{2\pi} d\phi\,n_{\rm a}(|\rvec-\lambda\Uvec|),
\end{eqnarray}
where $C = C_{\rm b}(\nu)$ is a frequency dependent normalisation factor such that $\int dU\,n_{\rm b}(U,\nu) = N_{\rm a}(N_{\rm a}-1)/2$. 
The frequency dependence of $n_{\rm b}$ arises due to the frequency dependence of the baselines ($U \propto \nu$).

\begin{figure}
\includegraphics[width= 8.5cm]{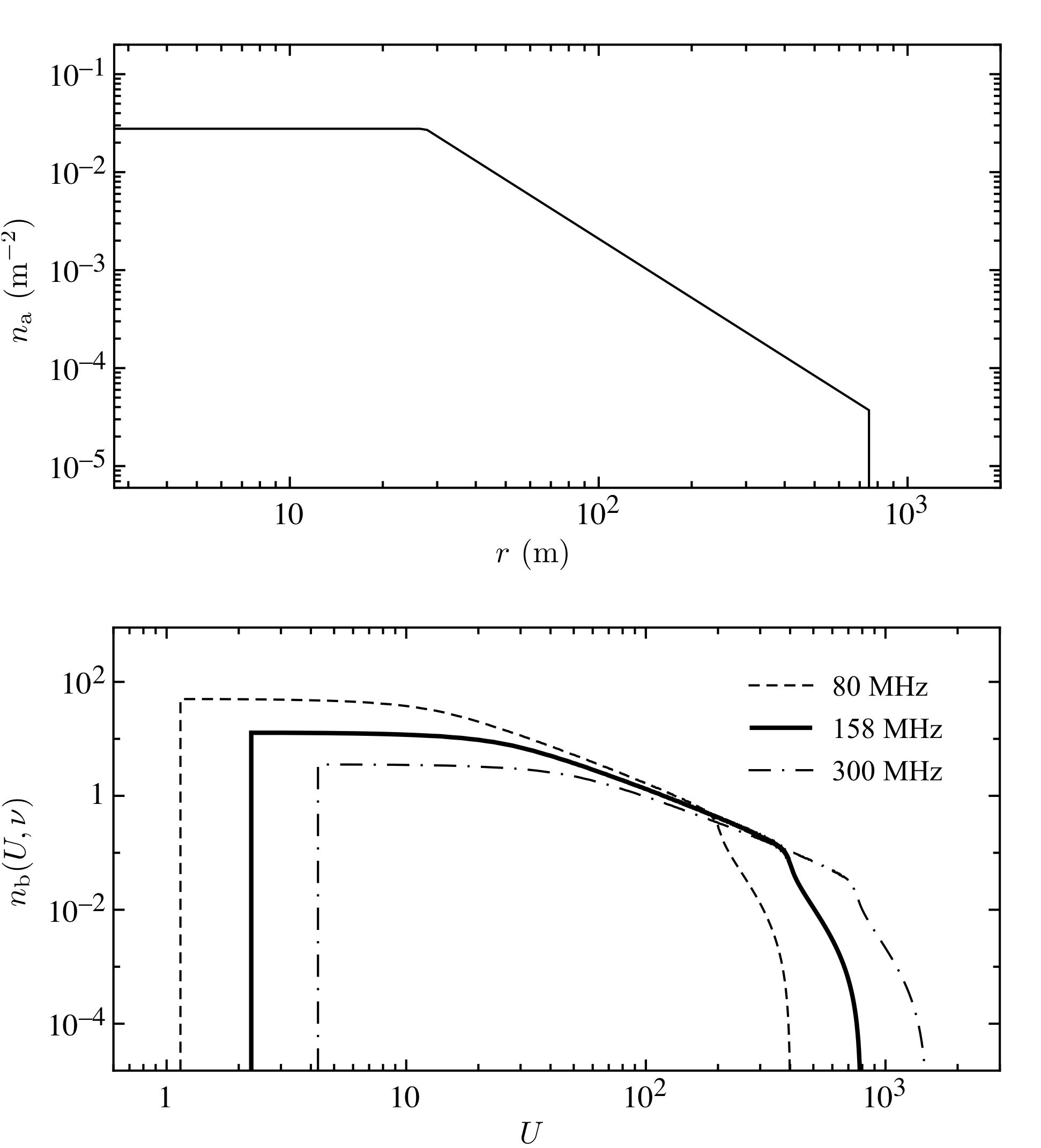} 
\caption{\textit{Top:} Antenna number density for a continuous $n_{\rm a} \propto r^{-2}$ distribution. \textit{Bottom:} Continuous baseline number density for $\nu = 80$, 158 and 300~MHz, corresponding to $z = 16.8$, 8 and 3.7 respectively.}
\label{fig:cont_dist}
\end{figure}

Figure~\ref{fig:cont_dist} shows the corresponding antenna number density for a continuous $n_{\rm a} \propto r^{-2}$ distribution, and the continuous baseline number density for $\nu = 80$, 158 and 300~MHz.

\subsection{Power spectrum measurement}
\label{App:Power spectrum measurement}

Calculations of the sensitivity to the 21-cm power spectrum for an interferometer have been presented by a number of authors \citep[][]{bharadwaj2001,bharadwaj2003,morales2005,bowman2006,mcquinn2006}. We follow the procedure outlined by \cite{mcquinn2006}, drawing on results from \cite{morales2005} and \cite{bowman2006} for the dependence of the array antenna density on radius, $n_{\rm a}(r)$.

As mentioned in Section~\ref{Fiducial array configuration}, the baseline number density $n_{\rm b}$ is a function of frequency. This dependency results in a smaller range of baseline values and denser sampling at lower frequencies compared with the coverage at higher frequencies (assuming an identical visibility regridding scale is used at each frequency). We ignore this dependency here since for bandwidths of $B \lsim\,\,32$~MHz, centred upon a frequency of interest with respect to the EOR, the relative scale difference in baseline for any antenna pair at each end of the frequency band is $0.8 \lsim\,\,U(\nu_{\rm min})/U(\nu_{\rm max}) = 1 - B/\nu_{\rm max} < 1$.

The uncertainty in a measurement of the power spectrum has two separate components, due to (i) the thermal noise of the instrument and (ii) sample variance. Here we introduce the power spectrum in the full Fourier representation and the first-order covariance matrices used to derive the relevant measurement errors, as well as discuss the relationship between instrumental and cosmological wave modes. Instrumental noise and sample variance are discussed in Section~\ref{App:Instrumental noise} and Section~\ref{Sample variance} respectively.

The wavevector $\kvec$ can be conveniently split into two orthogonal components, one parallel to the line of sight and the other transverse to it, such that $\kvec = \kvec_\parallel + \kvec_\perp$. The transverse wavenumber $k_\perp = |\kvec_\perp|$ is given by
\begin{eqnarray}
k_\perp \equiv \frac{2\pi}{d_\perp}  = \frac{2\pi}{D(z)\Delta\theta} = \frac{2\pi U}{D(z)}, \quad D(z) = \int^z_0 \frac{c\,dz^\prime}{H(z^\prime)},
\end{eqnarray}
where $d_\perp$ is the comoving dimension of the mode perpendicular to the line of sight,  $\Delta\theta$ is the angle subtended by $d_\perp$ on the sky (in radian) and $D(z)$ is the comoving distance to the point of emission at redshift $z$. The parallel wavenumber $k_\parallel = |\kvec_\parallel|$ is given by 
\begin{eqnarray}
\label{k_relations}
k_\parallel \equiv \frac{2\pi}{d_\parallel} = \frac{2\pi B}{\Delta D \Delta\nu}\,,\quad \Delta D \approx \frac{c(1+z)^2B}{\nu_{21} H(z)},
\end{eqnarray}
where $d_\parallel$ is the comoving dimension of the mode parallel to the line of sight, $\Delta D$ is the comoving depth of survey volume corresponding to the bandwidth $B$ and $\nu_{21}$ is the rest-frame frequency of 21-cm line. $dU = U_{\rm max}/N_{U}$ and $d\eta = 1/B$ with $\eta_{\rm max} =  N_{\rm ch}/B$.
\begin{eqnarray}
dk_\perp = \frac{2\pi dU}{D(z)}\,,\qquad dk_\parallel = \frac{2\pi}{\Delta D}.
\end{eqnarray}
Using these definitions we can write $P_{21}(\kvec) = (D^2 \Delta D/B)P_{21}(\Upsilonvec)$.

In the full $(u,v,\eta)$ Fourier representation, the measured visibility $V(\Uvec,\nu)$ is given by
\begin{eqnarray}
\label{VtoItilde}
\tilde{I}(\Upsilonvec) = \int d\nu\,V(\Uvec,\nu)\exp(2\pi i \nu\eta),
\end{eqnarray}
where the tilde indicates the two-dimensional Fourier transform that relates the specific intensity $I(\theta_x,\theta_y,\nu)$ in the image cube with the visibility $V(\Uvec,\nu)$ in the visibility cube.

The error in the 21-cm power spectrum \textit{per mode} of volume $d^3\Upsilonvec = du\,dv\,d\eta \approx A_{\rm e}/\lambda^2B$ is given by
\begin{eqnarray}
\label{Perror}
\delta P_{21}(\Upsilonvec) = \Big[C^{\rm N}(\Upsilonvec) + C^{\rm SV}(\Upsilonvec)\Big]d^3\Upsilonvec,
\end{eqnarray}
where $C^{\rm N}(\Upsilonvec)$ is the rms of instrumental noise and $C^{\rm SV}(\Upsilonvec)$ is the rms of the sample variance. The covariance of two measured $\tilde{I}(\Upsilonvec)$ values is
\begin{eqnarray}
\label{CXij}
C^{\rm X}_{ij} & \equiv & C^{\rm X}(\Upsilonvec_i,\Upsilonvec_j)\nonumber \\
&=& \Big\langle \tilde{I}^{\rm X}(\Upsilonvec_i) \Big[\tilde{I}^{\rm X}(\Upsilonvec_j)\Big]^* \Big\rangle,
\end{eqnarray}
where $C^{\rm X}(\Upsilonvec) = C^{\rm X}(\Upsilonvec,\Upsilonvec)$ and X represents the source of the error.

\subsubsection{Instrumental noise}
\label{App:Instrumental noise}

The rms noise fluctuation per visibility per frequency channel is given by \citep{rw} 
\begin{eqnarray}
\label{noiserms}
V^{\rm N}_{\rm rms}(\Uvec,\nu) = \frac{\partial T_{\rm b}}{\partial I_\nu}\frac{2k_{\rm B}T_{\rm sys}}{A_{\rm e}\sqrt{\Delta\nu t_{\Uvec}}}~{\rm K},
\end{eqnarray} 
where $T_{\rm sys}$ is the system temperature (K), $A_{\rm e}$ is the effective area of one antenna (m$^2$), $t_{\Uvec}$ is the integration time for that visibility (s), $\Delta\nu$ is the frequency bin width (Hz) and $k_{\rm B}$ is the Boltzmann constant. For the system temperature, we use the mean sky brightness due to DGSE in Equation~(\ref{TGb}). Instrumental noise is uncorrelated in both $\Uvec$ and $\nu$, i.e., $\langle V^{\rm N}_{\rm rms}(\Uvec_i,\nu_m)[V^{\rm N}_{\rm rms}(\Uvec_j,\nu_n)]^*\rangle \propto \delta_{ij}\delta_{mn}$. The average integration time $t_{\Uvec}$ that an array observes the visibility $\Uvec$ is given by
\begin{eqnarray}
t_{\Uvec} = \frac{A_{\rm e}}{\lambda^2}t_0 n_{\rm b}(U,\nu),
\end{eqnarray}
where $t_0$ is the total integration time and $n_{\rm b}(U,\nu)$ is the number density of baselines that can observe the visibility $\Uvec$.

Performing the discrete version of the transformation in Equation~(\ref{VtoItilde}) for the noise-only rms visibilities $V^{\rm N}_{\rm rms}$ in Equation~(\ref{noiserms}) gives
\begin{eqnarray}
\tilde{I}^{\rm N}(\Upsilonvec) = \sum_{m=1}^{N_{\rm ch}} V^{\rm N}_{\rm rms}(\Uvec,\nu_m)\exp(2\pi i \nu_m\eta) \Delta\nu,
\end{eqnarray}
where $N_{\rm ch} = B/\Delta\nu$ is the number of frequency channels, each centred on a frequency $\nu_j$. The covariance matrix for instrumental noise is found using Equation~(\ref{CXij}), and is
\begin{eqnarray}
\label{CNij_general}
C^{\rm N}_{ij} &=& \Big\langle \tilde{I}^{\rm N}(\Upsilonvec_i) \Big[\tilde{I}^{\rm N}(\Upsilonvec_j)\Big]^* \Big\rangle \nonumber \\
&=& \Bigg\langle \Bigg[\sum_{m=1}^{N_{\rm ch}} V^{\rm N}_{\rm rms}(\Uvec_i,\nu_m)\exp(2\pi i \nu_m\eta_i) \Delta\nu \Bigg]\nonumber \\
&\times & \Bigg[\sum_{n=1}^{N_{\rm ch}} V^{\rm N}_{\rm rms}(\Uvec_j,\nu_n)\exp(2\pi i \nu_n\eta_j) \Delta\nu \Bigg]^* \Bigg\rangle \nonumber \\
&=& (\Delta\nu)^2 \sum_{m,n=1}^{N_{\rm ch}} \Big\langle V^{\prime \rm N}_{\rm rms}(\Uvec_i,\nu_m) \Big[V^{\prime \rm N}_{\rm rms}(\Uvec_j,\nu_n)\Big]^*  \Big\rangle \nonumber \\
&=& (\Delta\nu)^2 \sum_{m=1}^{N_{\rm ch}} \Big\langle V^{\prime \rm N}_{\rm rms}(\Uvec_i,\nu_m) \Big[V^{\prime \rm N}_{\rm rms}(\Uvec_j,\nu_m)\Big]^* \Big\rangle \nonumber \\
&=& (\Delta\nu)^2 \sum_{m=1}^{N_{\rm ch}} \big|V^{\rm N}_{\rm rms}(\Uvec_i,\nu_m)\big|^2 \delta_{ij},
\end{eqnarray}
where $V^{\prime \rm N}_{\rm rms}(\Uvec,\nu) = V^{\rm N}_{\rm rms}(\Uvec,\nu)\exp(2\pi i \nu\eta)$. In the case of a contiguous bandwidth with $B \ll \nu_0$, $V^{\rm N}_{\rm rms}(\Uvec_i,\nu)$ is approximately the same in all frequency channels and the last expression in Equation~(\ref{CNij_general}) simplifies to
\begin{eqnarray}
\label{CNij}
C^{\rm N}_{ij} &=& N_{\rm ch} (\Delta\nu)^2 \big|V^{\rm N}_{\rm rms}(\Uvec_i,\nu)\big|^2 \delta_{ij} \nonumber \\
&=& \left( \frac{\lambda^2 T_{\rm sys}}{A_{\rm e}} \right)^2 \frac{B}{t_{\Uvec_i}} \delta_{ij} \nonumber \\
&=& \left(\frac{\lambda^2BT_{\rm sys}}{A_{\rm e}}\right)^2 \frac{\lambda^2}{BA_{\rm e}t_0n_{\rm b}(U_i,\nu_i)} \delta_{ij}.
\end{eqnarray}

\subsubsection{Sample variance}
\label{Sample variance}

The second component of uncertainty is due to sample variance within the finite volume of the survey. The covariance matrix for sample variance is given by
\begin{eqnarray}
\label{CSVij}
C^{\rm SV}_{ij} &=& \Big\langle \tilde{I}^{21}(\Upsilonvec_i) \Big[\tilde{I}^{21}(\Upsilonvec_j)\Big]^* \Big\rangle \nonumber \\
&=& \int d^3\Upsilonvec\, P_{21}(\Upsilonvec)\tilde{W}(\Upsilonvec_i - \Upsilonvec)\tilde{W}^*(\Upsilonvec_j - \Upsilonvec) \nonumber \\
& \approx & \frac{\lambda^2B}{A_{\rm e}} P_{21}(\Upsilonvec_i)\delta_{ij},
\end{eqnarray}
where $W$ is the instrumental window function given by the field of view and bandwidth of the array. The last line follows by performing the integral after noting that the Fourier transform of the window function, $\tilde{W}$ is non-zero only in a volume $du\,dv\,d\eta \approx A_{\rm e}/\lambda^2B$ and therefore $\tilde{W}$ is normalised by $(du\,dv\,d\eta)^{-1/2}$.\vspace{2mm}\\

Combining Equations~(\ref{Perror}), (\ref{CNij}) and (\ref{CSVij}), the resulting error in the 21-cm power spectrum \textit{per mode} is
\begin{eqnarray}
\delta P_{21}(\Upsilonvec) = \left(\frac{\lambda^2T_{\rm sys}}{A_{\rm e}}\right)^2 \frac{1}{t_0 n_{\rm b}(U,\nu)} + P_{21}(\Upsilonvec).
\end{eqnarray}
Written in terms of the cosmic wavevector,
\begin{eqnarray}
\delta P_{21}(\kvec) = \frac{T_{\rm sys}^2}{Bt_0} \frac{D^2 \Delta D}{n_{\rm b}(U,\nu)} \left(\frac{\lambda^2}{A_{\rm e}}\right)^2  + P_{21}(\kvec),
\end{eqnarray}
where $U = k_\perp D/2\pi$.

\subsection{Uniform gains}
\label{Uniform gains}

For a single baseline with uniform gains (equal to unity, therefore all $\gamma$'s vanish), Equations~(\ref{I_prime_b})--(\ref{U_prime_b}) simplify to
\begin{eqnarray}
\label{eq:Pprime1b}
I^{\prime}_{ij} &=& I_{ij} + \frac{1}{2}\delta_{+-}U_{ij} - \frac{1}{2}i\delta_{-+}V_{ij},\\
Q^{\prime}_{ij} &=& Q_{ij} + \frac{1}{2}\delta_{++}U_{ij} - \frac{1}{2}i\delta_{--}V_{ij},\\
U^{\prime}_{ij} &=& U_{ij} + \frac{1}{2}\delta_{+-}I_{ij} - \frac{1}{2}\delta_{++}Q_{ij}.
\end{eqnarray}
$\textbf{M}_{ij}$ has to be summed over all baselines and their conjugates however, which then gives the following expressions for the response in Stokes~$I$, $Q$ and $U$ (all of which are frequency and sky position dependent):
\begin{eqnarray}
\label{Iprime}
\bar{I}^{\prime} &=& \bar{I} + \frac{1}{4N_{\rm b}}\Bigg(\bar{U}\sum_{\forall b,b^*}\delta_{+-} - i\bar{V}\sum_{\forall b,b^*}\delta_{-+}\Bigg);\\
\label{Qprime}
\bar{Q}^{\prime} &=& \bar{Q} + \frac{1}{4N_{\rm b}}\Bigg(\bar{U}\sum_{\forall b,b^*}\delta_{++} - i\bar{V}\sum_{\forall b,b^*}\delta_{--}\Bigg);\\
\label{Uprime}
\bar{U}^{\prime} &=& \bar{U} + \frac{1}{4N_{\rm b}}\Bigg(\bar{I}\sum_{\forall b,b^*}\delta_{+-} - \bar{Q}\sum_{\forall b,b^*}\delta_{++}\Bigg),
\end{eqnarray}
where $N_{\rm b} = N(N-1)/2$ is the number of baselines and  $\bar{I} = \sum I_{ij}/(2N_{\rm b})$ etc. If we now assume that the sources have no circularly polarised component ($\bar{V} \sim 0$) and $\sum \Re[\Delta^{+}_{xx}] \approx \sum \Re[\Delta^{+}_{yy}]$ (this constraint only affects the Stokes~$I$ leakage into Stokes~$U$), where the $\Delta^{\pm}_{ab}$ terms are defined by
\begin{eqnarray}
\label{eq:Deltapm}
\Delta^{\pm}_{xx} &=& d_{i,xy} \pm d^*_{j,xy},\\
\Delta^{\pm}_{xy} &=& d_{i,xy} \pm d^*_{j,yx},\\
\Delta^{\pm}_{yx} &=& d_{i,yx} \pm d^*_{j,xy},\\
\Delta^{\pm}_{yy} &=& d_{i,yx} \pm d^*_{j,yx},
\end{eqnarray}
then Equations~(\ref{Iprime})--(\ref{Uprime}) simplify to
\begin{eqnarray}
\label{Iprime2}
\bar{I}^{\prime} &=& \bar{I},\\
\label{Qprime2}
\bar{Q}^{\prime} &=& \bar{Q} + \frac{\bar{U}}{2N_{\rm b}}\sum_{i}\sum_{j \neq i}\left[\Re(\Delta_{xx}^{+})+\Re(\Delta_{yy}^{+})\right],\\
\label{Uprime2}
\bar{U}^{\prime} &=& \bar{U} - \frac{\bar{Q}}{2N_{\rm b}}\sum_{i}\sum_{j \neq i}\left[\Re(\Delta_{xx}^{+})+\Re(\Delta_{yy}^{+})\right].
\end{eqnarray}
The linearly polarised components $Q$ and $U$ can be written
\begin{eqnarray}
\label{QU}
\left(
\begin{array}{c}
Q^{\prime}\\U^{\prime}
\end{array}\right) \approx
\left(
\begin{array}{cc}
1 & L_{QU}\\
L_{UQ} & 1
\end{array}\right)
\left(
\begin{array}{c}
Q\\U
\end{array}\right),
\end{eqnarray}
where $L_{QU}$ and $L_{UQ}$ are real, frequency-dependent Stokes leakage parameters. Equation~(\ref{QU}) has been used in the example in Figure~\ref{singlebeg}, which shows simulated examples of the CHECK Faraday dispersion function, $F(\phi)$, for a single baseline illustrating the effect of leakage between Stokes~$Q$ and $U$, noise and DC offsets. In general, however, the gain/leakage errors for each baseline will be independent, thereby complicating the analysis above.

\end{document}